\documentclass[aps, prd, twocolumn, showpacs, floatfix, letterpaper,
  nofootinbib, superscriptaddress,longbibliography]{revtex4-1}

\usepackage{graphicx}
\usepackage{epsfig}
\usepackage{bm}
\usepackage{bbold}
\usepackage{amsfonts}
\usepackage{pgf}
\usepackage{color}
\usepackage[T1]{fontenc}
\usepackage[latin1]{inputenc}
\usepackage{graphicx}
\usepackage[english]{babel}
\usepackage{amsmath}
\usepackage{amssymb}
\usepackage{amsfonts}
\usepackage{makecell}
\usepackage{hyperref}
\usepackage[draft,deletedmarkup=xout]{changes}
\usepackage{mathtools}
\usepackage{soul}
\usepackage{ulem}
\usepackage{relsize}

\definecolor{green}{rgb}{0.19,0.64,0.54}
\definecolor{blue}{rgb}{0,0,1}
\definecolor{reddish}{rgb}{0.65, 0.2, 0.2}
\definecolor{darkgreen}{rgb}{0.2,0.7,0.3}
\definecolor{darkblue}{rgb}{0.3,0.40,0.48}
\definecolor{gray}{rgb}{.8,.8,.8}

\hypersetup{,
  pdftitle={Coherent},
  pdfauthor={J. de Cabo, P. Malkiewicz and P. Peter},
  colorlinks=true,       % false: boxed links; true: colored links
  linkcolor=darkblue,          % color of internal links
  citecolor=darkgreen,        % color of links to bibliography
  filecolor=reddish,      % color of file links
  urlcolor=blue,           % color of external links
  linktocpage=true
}

\newcommand{\dd}{\mathrm{d}}
\newcommand{\ex}{\mathrm{e}}
\newcommand{\GN}{G_\textsc{n}}

\newcommand{\Hu}{\mathcal{H}}

\newcommand{\ud}{\mathrm{d}}

\newcommand{\be}{\begin{equation}}
\newcommand{\ee}{\end{equation}}
\newcommand{\ba}{\begin{eqnarray}} 
\newcommand{\ea}{\end{eqnarray}}
\newcommand{\N}{\textsc{f}}
\newcommand{\MS}{\textsc{c}}
\newcommand{\Ms}{\mathfrak{M}_\textsc{s}}
\newcommand{\Ns}{\mathfrak{N}_\textsc{s}}
\newcommand{\Ts}{\mathfrak{T}_\textsc{s}}
\newcommand{\Ks}{\mathfrak{K}_\textsc{s}}
\newcommand{\Ls}{\mathfrak{L}_\textsc{s}}

\newcommand{\Mq}{\mathfrak{M}_\textsc{q}}
\newcommand{\Nq}{\mathfrak{N}_\textsc{q}}
\newcommand{\Rq}{\mathfrak{R}_\textsc{q}}
\newcommand{\Tq}{\mathfrak{T}_\textsc{q}}
\newcommand{\Lq}{\mathfrak{L}_\textsc{q}}
\begin{document}

\title{Unitarily inequivalent quantum cosmological bouncing models}

\author{Jaime de Cabo Martin}
\email{jaime.decabomartin@ncbj.gov.pl}
\affiliation{National Centre for Nuclear Research, Pasteura 7, 02-093
Warszawa, Poland}

\author{Przemys{\l}aw Ma{\l}kiewicz}
\email{Przemyslaw.Malkiewicz@ncbj.gov.pl}
\affiliation{National Centre for Nuclear Research, Pasteura 7, 02-093
Warszawa, Poland}

\author{Patrick Peter}
\email{peter@iap.fr}
\affiliation{${\cal G}\mathbb{R}\varepsilon\mathbb{C}{\cal O}$ -- Institut
d'Astrophysique de Paris, CNRS \& Sorbonne Universit\'e, UMR 7095
98 bis boulevard Arago, 75014 Paris, France}
\affiliation{Centre for Theoretical Cosmology, Department of Applied
Mathematics and Theoretical Physics, University of Cambridge, Wilberforce
Road, Cambridge CB3 0WA, United Kingdom}

\begin{abstract}
By quantising the background as well as the perturbations in a simple
one fluid model, we show that there exists an ambiguity in the choice
of relevant variables, potentially leading to incompatible
observational physical predictions. In a classical or quantum
inflationary background, the exact same canonical transformations lead
to unique predictions, so the ambiguity we put forward demands a
semiclassical background with a sufficiently strong departure from classical
evolution. The latter condition happens to be satisfied in bouncing 
scenarios, which may thus be having predictability issues. Inflationary
models could evade such a problem because of the monotonic behavior of
their scale factor; they do, however, initiate from a singular state
which bouncing scenarios aim at solving.
\end{abstract}

\date{\today}
\maketitle

\section{Introduction}

Cosmological perturbations are usually studied on a classical or
semiclassical background in the framework of inflation
\cite{Peter:2013avv}.  Most models of bouncing alternatives are either
based on a classical background
\cite{Battefeld:2014uga,Brandenberger:2016vhg} or it is assumed that
the semiclassical approximation ensures similar behaviour for the
perturbations. The purpose of this paper is to show that there might
be some important caveat that should be taken into account as an
unsolved ambiguity can emerge in a quantum bouncing scenario.
It is worth mentioning that already in classical backgrounds,
the notion of the initial vacuum state depends on the choice
of perturbation variables for quantization as noted e.g. in
\cite{Grain:2019vnq}. Herein, we show that once the background is
quantised, the physical ambiguity gets much stronger and
concerns the dynamics of mode functions as well. A
similar point was considered in Refs.~\cite{Kiefer:2011cc,Chataignier:2020fap}
for an inflationary background, leading to a vanishingly small effect.

To illustrate our point, we examine a simple model based on canonical
quantization of general relativity (GR) in which the matter content is
represented by a perfect fluid with constant equation of state $w \in
[0,1[$. We first recall the classical model in its Hamiltonian
formulation both for the background (singular) universe and
the perturbations, before moving to a quantum approach aiming
at resolving the classical singularity.

The paper is organized as follows. Sec.~\ref{sec_Class} first
introduces the classical model consisting of general relativity
sourced by a simple constant equation-of-state fluid, for which
we define the Hamiltonian version of both the singular background
and the divergent perturbations. We then move, in
Sec.~\ref{sec_QuantModel}, to the quantum model by assuming
a very general quantization procedure allowing to account for
self-adjointness issues on the half line; the perturbations are
then treated in the usual (canonical) way. The ensuing quantum
dynamics is examined in Sec.~\ref{sec_QuantDyn} where the semiclassical
approximation for the background is found to yield two different,
and incompatible, equations of motion for the perturbation modes,
leading to a potential ambiguity in the observational predictions.
Our conclusions are followed by an appendix showing an explicit
example of quantization based on coherent states with definite
fiducial vectors.

\section{Classical model}\label{sec_Class}

In this section we provide the definition of the classical model
together with two different but physically equivalent
parametrizations. The physical phase space for the model is introduced
together with the physical Hamiltonian that generates its dynamics
with respect to an internal clock. The solution to the classical
dynamics is briefly discussed.

\subsection{Hamiltonian formalism}

We assume the universe to be spatially compact, $\mathcal{M}\simeq
\mathbb{R}\times\mathbb{T}^3$, with coordinate volume we note
$\mathcal{V}_0$ below. Its evolution is supposed to be driven by a
perfect fluid that satisfies a barotropic equation of state $p=w\rho$,
with $-\frac{1}{3}<w<1$. The fully canonical formalism for the
perturbed Friedmann universe that can be easily adapted to the present
case can be found in \cite{Malkiewicz:2018ohk}: we start from the
Einstein-Hilbert-Schutz action \cite{Schutz:1970my,Schutz:1971ac}
\begin{align}\begin{split}
\mathcal{S}_\textsc{ehs}=\underbrace{\frac{1}{16\pi\GN}\int \dd^4 x
  \sqrt{-g} R}_{\mathcal{S}_\textsc{eh}} +\underbrace{\int \dd^4 x
  \sqrt{-g} P(w,\phi)}_{\mathcal{S}_\textsc{s}},\end{split}
\end{align}
where $P=w\rho$ is the pressure of the cosmic fluid while $\phi$
defines its flow. The action $\mathcal{S}_\textsc{ehs}$ is first
expanded to second order around the flat Friedmann universe.  Next the
Hamiltonian description is obtained in which the truly physical
degrees of freedom are identified and the remaining ones removed.

Let us consider the usual Einstein-Hilbert action
$\mathcal{S}_\textsc{eh}$ at zeroth order, omitting the integrated
term
\begin{equation}
\frac{1}{16\pi\GN} \int \dd^4 x \sqrt{-g} R = -
\frac{1}{2\kappa}\int \dd \tau N a^3 \underbrace{\int
  \sqrt{\gamma}\dd^3 x}_{\mathcal{V}_0} \underbracket{\frac{6
    \dot{a}^2}{a^2 N^2}}_{R},
\end{equation}
in which we used the background isotropic and homogeneous flat
Friedmann-Lema\^{\i}tre-Robertson-Walker (FLRW) metric
\begin{equation}
\dd s^2 = -N^2(\tau)\dd\tau^2 + a^2(\tau) \gamma_{ij}\dd x^i \dd x^j,
\label{FLRW}
\end{equation}
a dot meaning a derivative with respect to the coordinate time $\tau$,
later to be identified with the fluid clock variable.  Written as
$\mathcal{S}_\textsc{eh} = \int L^{(0)}(a, \dot{a})\dd\tau$, with
Lagrangian $L^{(0)} = 3\mathcal{V}_0 a \dot{a}^2/(N\kappa)$, this
yields the canonically conjugate momentum $p_a = \partial
L^{(0)}/\partial \dot{a} = 6\mathcal{V}_0 a \dot{a}/(\kappa N)$, and
the gravitational Hamiltonian at zeroth order $H^{(0)}_\textsc{g}$
reads
\begin{equation}
H^{(0)}_\textsc{g} = -\frac{\kappa N}{12 \mathcal{V}_0 a} p_a^2,
\end{equation}
which can also be expressed in terms of the canonical variables,
\begin{equation}
q=\frac{4\sqrt{6}}{3(1-w)\sqrt{1+w}}a^{\frac{3}{2}(1-w)}
\equiv \gamma a^{\frac{3}{2}(1-w)},
\label{defGamma}
\end{equation}
thereby defining the constant $\gamma$, and 
\begin{equation}
p=\frac{\sqrt{6(1+w)}}{2\kappa_0} a^{\frac{3}{2}(1+w)}H,
\end{equation}
where $H=\dot{a}/(Na)$ is the Hubble rate and
$\kappa_0=\kappa/\mathcal{V}_0$. The Hamiltonian $H^{(0)}_\textsc{g}$
reads
\begin{equation}
H^{(0)}_\textsc{g}= -\frac{2\kappa_0 N}{(1+w) a^{3w}} p^2 = -2\kappa_0
p^2,
\end{equation}
where in the last equality, we made the choice of the lapse
$N=(1+w)a^{3w}$. It can be shown that for this particular choice of
the lapse the matter Hamiltonian $H^{(0)}_\textsc{m}$ obtained from
the Schutz action $\mathcal{S}_\textsc{s}$ equals the cosmic fluid
conjugate momentum (see, e.g. \cite{Bergeron:2013ika} for details).
Therefore, the total Hamiltonian generates a uniform motion in the
cosmic fluid variable. It is a standard procedure at this point to
promote the cosmic fluid variable to the role of internal clock while
removing it and its conjugate momentum from the phase space.  The
physical Hamiltonian that generates the dynamics of the background
geometry with respect to the fluid variable is thus simply
$H^{(0)}_\textsc{g}$. However, we find it convenient to inverse the
direction of time with respect to the fluid variable in order to have
the positive physical Hamiltonian,
\begin{equation}
H^{(0)}=-H^{(0)}_\textsc{g}=2\kappa_0 p^2.
\label{h0}
\end{equation}
We shall denote the internal clock by $\tau$ and assume it coincides
with the FLRW time set in \eqref{FLRW}~\cite{Peter:2013avv}. It can be
shown that the Hamiltonian $H^{(0)}=(1+w)E_\text{f}|_{a=1}$ equals
$(1+w)$ times the energy of the fluid contained in the universe when
$a=1$ (we choose a dimensionless scale factor, so that the canonical
variable $q$ is also dimensionless).

After identification of the truly physical degrees of freedom also at
linear order, we write the full Hamiltonian $H_\text{full}$ as
\begin{equation}
H_\text{full}=H^{(0)}-\sum_{\bm{k}}H^{(2)}_{\bm{k}},
\label{h1}
\end{equation}
where the second-order Hamiltonian $H^{(2)}_{\bm{k}}$, depending only
on the discrete (recall the Universe considered is compact) wavevector
$\bm{k}$, reads
\begin{equation}
H^{(2)}_{\bm{k}}=
\frac{1}{2}|{\pi_{\phi,{\bm{k}}}}|^2 +
\frac12 w (1+w)^2 
\left(\frac{{q}}{\gamma}\right)^{\frac{4(3w-1)}{3(1-w)}}
k^2|{\phi}_{\bm{k}}|^2,
\label{HAMpert}
\end{equation}
with $\gamma$ defined in \eqref{defGamma} above.  The Fourier
component $\phi_{\bm{k}}$ of the perturbation field is a combination
of the fluid perturbation\footnote{The background fluid time $\tau$ is
  actually a combination of the fluid variable and its momentum,
  $(1+w) \tau=\phi |p_\phi|^{-1/w}$. For more details, see
  e.g. \cite{Malkiewicz:2018ohk}.}  $\delta\phi_{\bm{k}}$ and the
intrinsic curvature perturbation $\delta R_{\bm{k}}$, namely
\cite{Malkiewicz:2018ohk}
\begin{equation}
\phi_{\bm{k}}=\frac{p^{\frac{1-w}{1+w}}\delta\phi_{\bm{k}}}{\sqrt{2w(1+w)\kappa_0}}
+ \sqrt{\frac{3}{w\kappa_0}} \frac{a^{-\frac{3w-7}{2}}}{4 k^2}\delta
R_{\bm{k}},
\end{equation}
with $k\equiv |{\bm{k}}|$ the amplitude of the wavevector; note that
since the FLRW background \eqref{FLRW} is isotropic, it is expected,
as usual, that the initial conditions, and therefore the solutions of
the perturbation evolution equation should depend only on the
amplitude $k$ and not on its direction ${\bm{k}}/k$.  Given our
conventions, the physical dimensions are $[\phi_{\bm{k}}]=\sqrt{M}L$
and $[\pi_{\phi,{\bm{k}}}]=\sqrt{M}$. The Poisson bracket reads
$\{\phi_{\bm{k}_1},\pi_{\phi,-\bm{k}_2}\}=
\delta_{{\bm{k}}_1,{\bm{k}}_2}$. The equation of motion expressed in
the conformal time $\eta$ defined below [see Eq.~\eqref{eta}], is
found to read
\begin{equation}
\phi_{\bm{k}}^{\prime\prime}+\left(
\frac{{q}}{\gamma}\right)^{\frac{4(3w-1)}{3(1-w)}}
w(1+w)^2k^2\phi_{\bm{k}}=0.
\end{equation}
It shows that for radiation, i.e. for $w=\frac13$, the dynamics of
$\phi_{\bm{k}}$ becomes decoupled from the dynamical background.

There exists infinitely many parametrizations of the reduced phase
space of perturbations and the initial parametrization
$(\phi,\pi_\phi)$  is just one
example. We shall call it the fluid-parameterization,
as the
relevant time for that description is $\tau$, which stems
from the fluid.  As another example, let us consider a pair of canonical
fields $(v,\pi_v)$ that is commonly used for solving the dynamics of
scalar perturbations
\begin{equation}\label{ntoms}
v_{\bm{k}}=Z \phi_{\bm{k}}, \qquad \pi_{v,{\bm{k}}}=Z^{-1}\pi_{\phi,{\bm{k}}}+
\frac{\dot{Z}}{Z^2}\phi_{\bm{k}},
\end{equation}
where
\begin{equation}
Z(\tau)=\sqrt{1+w}\left( \frac{q}{\gamma} \right)^{\frac{3w-1}{3(1-w)}}
\label{Zdef}
\end{equation}
and $\tau$, as mentioned above, denotes the internal fluid time.  Note
that in the comoving gauge $\delta\phi_{\bm{k}}=0$, and thus
$v_{\bm{k}}=-\sqrt{\frac{3(1+w)}{16w\kappa_0}} a \Psi_{\bm{k}}$, where
$\Psi_{\bm{k}}=- a^2 \delta R_{\bm{k}}/k^2$ is the comoving curvature.
We easily obtain the second-order Hamiltonian $H^{(2)}_{\bm{k}}$ in
terms of $(v,\pi_v)$, namely
\begin{equation}
H^{(2)}_{\bm{k}} = \frac12 Z^2 \left\{ |\pi_{v,{\bm{k}}}|^2 
+ \left[ w k^2 - \mathcal{V}_\text{cl}(\tau) \right]|v_{\bm{k}}|^2
\right\},
\label{H2k}
\end{equation}
with the potential $\mathcal{V}_\text{cl}$ defined through
\begin{equation}
\mathcal{V}_\text{cl} = \frac{1}{Z^4} \left[ \frac{\ddot{Z}}{Z} -2 \left( 
\frac{\dot{Z}}{Z} \right)^2 \right]
\label{calV}
\end{equation}
which can be written explicitly in terms of the background canonical
variables $q$ and $p$ as
\begin{widetext}
%\begin{equation}\label{HamMS}
%H^{(2)}_{\bm{k}} =
%\frac{1+w}{2}\left(\frac{{q}}{\gamma}\right)^{\frac{2(3w-1)}{3(1-w)}}
%\left\{|\pi_{v,k}|^2+
%\left[ w k^2 - \frac{8\mathfrak{g}^2}{9{q}^2} 
%\left(\frac{{q}}{\gamma}\right)^{\frac{4(1-3w)}{3(1-w)}}
%\frac{(1-3w){p}^2}{(1+w)^2(1-w)^2}\right]
%|v_k|^2\right\}.
%\end{equation}
\begin{equation}\label{HamMS}
H^{(2)}_{\bm{k}} = \frac12 (1+w) \left( \frac{{q}}{\gamma}
\right)^{\frac{2(3w-1)}{3(1-w)}} \left\{|\pi_{v,{\bm{k}}}|^2+ \left[ w
  k^2 - \frac{8}{9{q}^2}
  \left(\frac{{q}}{\gamma}\right)^{\frac{4(1-3w)}{3(1-w)}}
  \frac{(2\kappa_0)^2 (1-3w){p}^2}{(1+w)^2(1-w)^2}\right]
|v_{\bm{k}}|^2\right\},
\end{equation}
\end{widetext}
where we used the background equations of motion by assuming
$p\to$~const.; we shall call the set of variables
$(v_{\bm{k}},\pi_{v,{\bm{k}}})$ the conformal parametrization, 
as it involves naturally the conformal time.
It differs from the fluid parametrization \eqref{HAMpert} by the
nontrivial coefficient standing in front of the entire expression as
well as the frequency that now depends on both background variables,
${q}$ and ${p}$.

The coefficient in front of the Hamiltonian \eqref{HamMS} can be
removed by switching to the internal conformal time $\eta$
\cite{Malkiewicz:2015fqa}
\begin{equation}
\dd\eta=Z^2 \dd\tau = (1+w)\left(
\frac{q}{\gamma}\right)^{\frac{2(3w-1)}{3(1-w)}}\dd\tau,
\label{eta}
\end{equation}
in terms of which the potential \eqref{calV} takes the simpler and
usual form $\mathcal{V}_\text{cl} = Z''/Z$, where a prime means a
derivative with respect to the conformal time $\eta$ ($Z' \equiv \dd
Z/\dd\eta$).  The second-order Hamiltonian $Z^{-2} H^{(2)}_{\bm{k}}$
is then found to generate
\begin{equation}\label{vVcl}
v''_{\bm{k}} + \left[ w k^2 - \frac{8}{9 q^2 Z^4} 
\frac{(2\kappa_0)^2 (1-3w)}{(1-w)^2} p^2 \right] v_{\bm{k}}=0,
\end{equation}
which can be written in the usual Mukhanov-Sasaki form
\begin{equation}\label{eomP}
v''_{\bm{k}} + \left[ w k^2 - \mathcal{V}_\text{cl} (\eta) \right] =
v''_{\bm{k}} + \left( w k^2 - \frac{z''}{z}\right) v_{\bm{k}}=0,
\end{equation}
thereby identifying the classical potential
\begin{equation}
 \mathcal{V}_\text{cl} (\eta) = \frac{8}{9 q^2 Z^4} 
\frac{(2\kappa_0)^2 (1-3w)}{(1-w)^2} p^2 = \frac{z''}{z},
\label{Vclass}
\end{equation}
where the last equality is obtained by applying the classical
equations of motion Eq.~\eqref{eomB} below and we have
used the generic function $z$, as there are in fact two
different and equivalent choices that can be made, namely
$z_1=q^{r_1}$ and $z_2=q^{r_2}$, with
\begin{equation}
r_1 = \frac{3w-1}{3(1-w)} \qquad \hbox{and} \qquad 
r_2 = \frac{2}{3(1-w)}.
\label{r1r2}
\end{equation}
These two power laws stem from the fact that although what enters into
\eqref{H2k} is $Z''/Z$, with $Z\propto z_1$, one can then just as well
choose the second solution of $z''/z = Z''/Z$, namely $z_2\propto Z
\int \dd\eta/Z^2 = Z \int \dd \tau = Z \tau$ which, taking the
background solution $q \propto \tau$ [see Eq.~\eqref{qpSol} below]
yields $z_2 \propto Z q = z_1 q = q^{r_1+1}$, as indeed one has $r_2 =
r_1 +1$.

The internal conformal time provides a convenient form of the equation
of motion for perturbations. We shall, however, quantize the dynamics
of both the background and the perturbations reduced with respect to a
unique internal time, the internal fluid time. The term $z''/z$ is
usually referred to as the potential for the perturbations, as
Eq.~\eqref{eomP} is mathematically identical to a time-independent
Schr\"odinger equation in such a potential \cite{Martin:2003bp}.  The
potential
\begin{equation}\label{gravpot}
\mathcal{V}_\text{cl} = \frac{z''}{z}=\frac{1-3w}{2}
  \Hu^2,
\end{equation}
has the clear physical meaning of the conformal Hubble rate $\Hu$
squared ($w<\frac13$). Therefore, the conformal Hubble rate determines
the coordinate scale at which the amplification of perturbations
starts to take place.

\subsection{Background solution}

The background Hamilton equations,
\begin{equation}\label{eomB}
\frac{\dd q}{\dd \tau}=4\kappa_0 p\quad
\text{and}
\quad \frac{\dd p}{\dd \tau}=0,\end{equation}
have the solution
\begin{equation}
q(\tau)=\sqrt{8\kappa_0 H^{(0)}}(\tau-\tau_\text{s})\quad
\text{and}
\quad p(\tau)=\sqrt{\frac{H^{(0)}}{2\kappa_0}},
\label{qpSol}
\end{equation}
where $H^{(0)}$ is the value of the zeroth-order Hamiltonian, a
constant by virtue of its definition \eqref{h0} and the equation of
motion \eqref{eomB}. The phase space trajectories that either
terminate at or emerge from the singularity at time $\tau_\text{s}$
are straight lines in phase space $\{ q, p\}$ with constant
$p$~\cite{Malkiewicz:2015fqa}, shown as straight lines in
Fig.~\ref{fig:Qtraj}.  Note that in order to assign the correct
trajectory to the background universe, one needs to know the value of
the energy of the fluid in the whole universe when $a=1$. This value
can be determined only when one knows the size of the universe, size
which can be fixed by demanding that the volume of the observable
patch be a given ratio (less than unity) of the size of the full
universe.

Anticipating the quantum solution \eqref{solsem}, we write the
classical solution as $q = q_\textsc{b} \omega \tau$ (setting the
singularity time to $\tau_\textsc{s} \to 0$), and therefore $p =
q_\textsc{b} \omega /(4\kappa_0)$. Eq.~\eqref{eta} with this solution
permits to integrate explicitly for the conformal time $\eta$, also
assuming $\eta\to0$ for $\tau\to 0$.  One finds the classical
conformal time to read
\begin{equation}
\eta = \frac{1+w}{2 r_1 +1} \left( \frac{q_\textsc{b} \omega}{\gamma}
\right)^{2 r_1} \tau^{2 r_1 +1},
\label{etaCl}
\end{equation}
which is straightforwardly inverted to yield $\tau (\eta)$, and finally
\begin{equation}
q(\eta) = q_\textsc{b} \omega \left[ \frac{2 r_1 +1}{1+w} \left(
  \frac{q_\textsc{b} \omega}{\gamma} \right)^{-2 r_1} \eta
  \right]^{1/(2r_1 +1)} \propto \eta^{\frac{3(1-w)}{1+3w}},
\label{qeta}
\end{equation}
and the classical potential then reads
\begin{equation}\label{Vcleta}
\mathcal{V}_\text{cl} = \frac{(q^{r_1})''}{q^{r_1}} =
\frac{(q^{r_2})''}{q^{r_2}} =\frac{2(1-3w)}{(1+3w)^2 \eta^2},
\end{equation}
as usual for a background dominated by a perfect fluid with constant
equation of state.

\subsection{Solution for perturbations}

The two parametrizations described above, $(\phi,\pi_{\phi})$ and
$(v,\pi_{v})$, are physically equivalent and therefore it is
sufficient to consider just one of them, e.g., the conformal one, in
order to determine the dynamics of perturbations. Using the definition
\eqref{eta} to derive the power-law behaviour of $q(\eta)$ in
\eqref{qpSol}, the potential $z''/z$ in Eq.~\eqref{eomP} is found to
yield the specific form \eqref{Vcleta} (independently of the choice
$z=z_1$ or $z=z_2$), so that the classical evolution of perturbation
modes is
\begin{equation}
\frac{\dd^2 v_{\bm{k}}}{\dd\eta^2}+\left[ w
  k^2-\frac{2(1-3w)}{(1+3w)^2\eta^2}\right] v_{\bm{k}}=0.
\label{Vcl}
\end{equation}
Clearly, the potential $\mathcal{V}_\text{cl} \propto \eta^{-2}$ is
singular at this point too. The solution can be expressed in terms of
Hankel functions, namely
\begin{equation}
v_{\bm{k}}(\eta)=\sqrt{\eta} \left[c_1({\bm{k}})
H_{\nu}^{(1)}\left( \sqrt{w}k\eta \right) +
c_2({\bm{k}}) H_{\nu}^{(2)} \left( \sqrt{w}k\eta \right)
\right],
\label{vHankel}
\end{equation}
where $\nu = \frac{3(1-w)}{2(3w+1)}$ and $c_1({\bm{k}})$,
$c_2({\bm{k}})$ are constants depending on the comoving wavevector
${\bm{k}}$ through the initial conditions; for isotropic initial
conditions as those used for quantum vacuum fluctuation, they can
depend only on the amplitude $k$ and not on the direction
${\bm{k}}/k$. The solution is finite but discontinuous at
$\eta=0$. Therefore, the comoving curvature $\Psi_{\bm{k}}\propto
v_{\bm{k}}/a$ in general blows up at $\eta=0$ where the scale factor
reaches the singularity $a\to0$; see Ref.~\cite{Peter:2001fy} for a
full treatment of the relevant cases.

\section{Full quantum model}
\label{sec_QuantModel}

In the present section we quantize the Hamiltonian \eqref{h1} in the
two parametrizations we introduced above. Next we apply some
approximations in order to integrate the dynamics. We find that the
two classically equivalent parametrizations lead to two unitarily
inequivalent quantum theories. This dependence on parametrization is a
natural consequence of the nonlinearity of the theory of
gravity. Recall that Dirac's ``Poisson bracket $\rightarrow$
commutator" quantization rule \cite{Dirac:1925jy} works only for
simplest observables. It is well-known that there exists no
quantization of any given classical system that is an isomorphism
between the Poisson and commutator algebras (there is actually one
known exception that nevertheless is irrelevant in the present
context, see \cite{Gotay:1996pmr} for an exhaustive review). As a
result, a quantized observable is in general unitarily inequivalent if
its quantization is made with a different choice of basic
observables. Note that this ``obstruction" is absent when quantum
perturbations evolve linearly in classical backgrounds as in the
latter case all the possible sets of basic variables are related by
linear transformations that enjoy unique unitary representations
consistent with Dirac's rule.

The phase space for the cosmological background is the half-plane
rather than the full plane and hence the usual canonical quantization
rules seem to be inadequate. There exist many quantization methods
(see, e.g. \cite{Ali:2004ft} for a comprehensive review) some of which
one could find more suitable in the present context. In order to
account for this issue, we introduce a family of quantum models, all
of which in correspondence with the underlying classical model. They
are given by a set of free parameters that can be computed, for
instance, in the framework of the so-called affine quantization (see
\cite{Gazeau:2015nkc} for details) that has been proposed for a
consistent quantum gravity programme
\cite{Klauder:2001bp,Klauder:2015ifa}; we briefly recap what is
relevant for our purposes of this method in Appendix \ref{A}.  This
approach enables us to free ourselves from a particular quantization
method and thus to emphasise the universal character of the
quantization ambiguity that we study below.

We then introduce a semiclassical approximation to the quantum
dynamics of the background geometry, which is a standard step in such
models. It should be noted that any ambiguous effect such as the one
we obtain here at a semiclassical level may only be enhanced if a
fully quantum description of the background were to be used. We
carefully construct the semiclassical description with the use of
coherent states.

\subsection{Background}

Given the existing ambiguities due to factor ordering when going from
classical to quantum, we propose the following set of operators to
replace the Hamiltonian \eqref{h0}:
\begin{equation}\label{qh0}
H^{(0)}\mapsto \widehat{H}^{(0)}=2\kappa_0\left(\widehat{P}^2+
{\hbar^2\mathfrak{c}_{0}}\widehat{Q}^{-2}\right),
\end{equation}
where $\mathfrak{c}_{0}\geqslant 0$ is a free parameter.  The value
$\mathfrak{c}_{0}=0$ corresponds to the ``canonical quantization",
whereas the values $\mathfrak{c}_{0}>0$ can be justified in various
ways, for instance by using the affine group as the symmetry of
quantization \cite{Malkiewicz:2019azw,Bergeron:2013ika}. In the latter
case, the repulsive potential $\propto \widehat{Q}^{-2}$, of quantum
geometric origin, prevents the universe from reaching the singular
point $q=0$ by reversing its motion from contraction to expansion. If
$\mathfrak{c}_{0}\geqslant \frac{3}{4}$, then
$\widehat{H}^{(0)}$ is essentially self-adjoint and no boundary
condition needs be imposed at $\widehat{Q}=0$ to ensure a unique and
unitary dynamics. The only way to determine the right value of the
parameter $\mathfrak{c}_{0}$ is to compare the predictions of the
model with the actual observations of the Universe.

We will need quantum operators to replace other zeroth-order
quantities, for those appear in the Hamiltonians relevant for
describing perturbations \eqref{HAMpert} and \eqref{HamMS}.  We
propose the following replacements
\begin{subequations}
\label{rules}
\begin{align}
q^\alpha & \mapsto \mathfrak{l}(\alpha) \widehat{Q}^\alpha, \label{qalpha}\\
q^{\alpha}p^2&\mapsto{\mathfrak{a}(\alpha)}\widehat{Q}^{\alpha}
\widehat{P}^2+i\hbar{\mathfrak{b}(\alpha)}\widehat{Q}^{\alpha-1}
\widehat{P}+{\hbar^2\mathfrak{c}(\alpha)}\widehat{Q}^{\alpha-2},\label{qalphap2}
\end{align}
\end{subequations}
% \begin{equation}\begin{split}
% q^{\alpha}&\mapsto{\mathfrak{l}(\alpha)}
% \widehat{Q}^{\alpha},\\
% q^{\alpha}p^2&\mapsto{\mathfrak{a}(\alpha)}\widehat{Q}^{\alpha}
% \widehat{P}^2+i{\mathfrak{b}(\alpha)}\widehat{Q}^{\alpha-1}
% \widehat{P}+{\mathfrak{c}(\alpha)}\widehat{Q}^{\alpha-2},
% \end{split}
% \label{ClassQ}
% \end{equation}
where $\widehat{Q}$ and $\widehat{P}$ are the `position' and
`momentum' operators on the half-line, satisfying the usual
commutation relation $[\widehat{Q},\widehat{P}] = i\hbar$, and
therefore $[\widehat{Q}^\alpha,\widehat{P}] = i\hbar\alpha
\widehat{Q}^{\alpha-1}$, so that ${\mathfrak{b}(\alpha})=-\alpha
        {\mathfrak{a}(\alpha)}$ in order to ensure that the
        second-line operator \eqref{qalphap2} is symmetric, i.e. so that it reads
\begin{equation}
q^{\alpha}p^2\mapsto \mathfrak{a}(\alpha)
\widehat{P}\widehat{Q}^{\alpha}\widehat{P} + \hbar^2\mathfrak{c}(\alpha)
\widehat{Q}^{\alpha-2};
\label{ClassQsym}
\end{equation}
the power-depending numbers $\mathfrak{l}(\alpha)$,
$\mathfrak{a}(\alpha)$ and $\mathfrak{c}(\alpha)$ are assumed
positive and  dimensionless.

\subsection{Perturbations}

The canonical perturbation variables of the fluid parametrization
satisfy the reality condition $\phi^*_{\bm{k}}=\phi_{-\bm{k}}$ and
$\pi^*_{\phi,\bm{k}}=\pi_{\phi,-\bm{k}}$ and it is possible to promote
their real and imaginary parts to canonical operators in
$L^2[\mathbb{R}^2,\frac{i}{2}\dd \phi_{\bm{k}}\dd\phi^*_{\bm{k}}]$ for
each direction $\bm{k}$. It is, however, more convenient to work with
the Fock representation,
\begin{equation}
  \phi_{\bm{k}}\mapsto\widehat{\phi}_{\bm{k}} =
  \sqrt{\frac{\hbar}{2}}\left[
    a_{\bm{k}}\phi^*_{k}(\tau)+{a}^{\dagger}_{-\bm{k}}\phi_{k}(\tau)\right],
\end{equation}
where the time-dependent mode functions ${\phi}_k (\eta)$ are assumed
to be isotropic and $a_{\bm{k}}$ and $a^\dagger_{\bm{k}}$ are fixed
annihilation and creation operators that satisfy
$[{a}_{\bm{k}_1},{a}^{\dagger}_{\bm{k}_2}]=\delta_{\bm{k}_1,\bm{k}_2}$
(recall the compactness of space implies discrete eigenvectors
$\bm{k}$).  As shown later, it follows that the mode functions must
satisfy a suitable normalization condition.  Note that the whole
evolution of the operators $\widehat{\phi}_{\bm{k}}$ and
$\widehat{\pi}_{\phi,\bm{k}}$ in the Heisenberg picture is encoded
into the mode functions.
 
Combining the background quantization with the quantization of
perturbations, using the definition \eqref{r1r2} of the classical
power laws, yields the quantized Hamiltonian \eqref{HAMpert} in the
fluid parametrization (henceforth dubbed \N-parametrization)
\begin{equation}\label{qham}
  \widehat{H}^{(2)}_{\bm{k}} =
  \frac{1}{2}|\widehat{\pi}_{\phi,{\bm{k}}}|^2 + \frac{\Lq}{2}
  w(1+w)^2 \left(
  \frac{\widehat{Q}}{\gamma}\right)^{4 r_1}
  k^2|\widehat{\phi}_{\bm{k}}|^2,
\end{equation}
where $\Lq=\mathfrak{l}(4 r_1)$ is a free parameter of the
quantization.

We repeat the same quantization for the conformal parametrization (\MS-parametrization in what follows),
\begin{equation}
  v_{\bm{k}} \mapsto \widehat{v}_{\bm{k}} =
 \sqrt{\frac{\hbar}{2}}\left[a_{\bm{k}} \bar{v}_k(\tau)+
  {a}^{\dagger}_{-\bm{k}} v_k(\tau)\right],
\end{equation}
and obtain the quantum \MS{} Hamiltonian derived from \eqref{HamMS} as
\begin{equation}
    \widehat{H}^{(2)}_{\bm{k}} = \frac12(1+w)
    \left(\frac{\widehat{Q}}{\gamma}\right)^{2 r_1}\Mq
    \, H^{(2)}_{{\bm{k}},\text{eff}},
    \label{H2kQ}
\end{equation}
with
\begin{widetext}
\begin{equation}\begin{split}
    \widehat{H}^{(2)}_{{\bm{k}},\text{eff}} =
    |\widehat{\pi}_{v,{\bm{k}}}|^2+\left[ wk^2-
      \frac{8\Mq^{-1}}{9\widehat{Q}^2}\frac{(2\kappa_0)^2(1-3w)}{(1-w)^2(1+w)^2}
      \left(\frac{\widehat{Q}}{\gamma}\right)^{-4r_1}
    \left(\Nq\widehat{P}^2+i\hbar\Rq \widehat{Q}^{-1}\widehat{P}+
        \hbar^2\Tq \widehat{Q}^{-2}\right)\right]
        |\widehat{v}_{\bm{k}}|^2,
\end{split}
\end{equation}
\end{widetext}
where $\Mq=\mathfrak{l}(2r_1)$, $\Nq=\mathfrak{a}(-2r_2)$,
$\Rq=\mathfrak{b}(-2r_2)=2 r_2 \Nq$ and $\Tq=\mathfrak{c}(-2r_2)$
are free parameters in the quantization map. Note that there are more
free parameters and hence more quantization ambiguities in the \MS{}
parametrization.

\section{Quantum dynamics} \label{sec_QuantDyn}

A general approach to solving the dynamics of quantum perturbations in
quantum spacetime was recently given in \cite{Malkiewicz:2020fvy}.  In
what follows, we assume the full state vector to be a product of
background and perturbation states, i.e.,
\begin{equation}
|\psi (\tau)\rangle=|\psi_\textsc{b}(\tau)\rangle\cdot
|\psi_\textsc{p}(\tau)\rangle.
\label{psiBP}
\end{equation}
The canonical formalism for cosmological perturbations has been
developed under the assumption that the perturbations do not backreact
on the background spacetime. Therefore, the dynamics of
$|\psi_\textsc{b}(\tau)\rangle$ should be determined independently of
the state $|\psi_\textsc{p}(\tau)\rangle$.

\subsection{Background semiclassical solution}

\begin{figure}
 \includegraphics[width=0.4\textwidth]{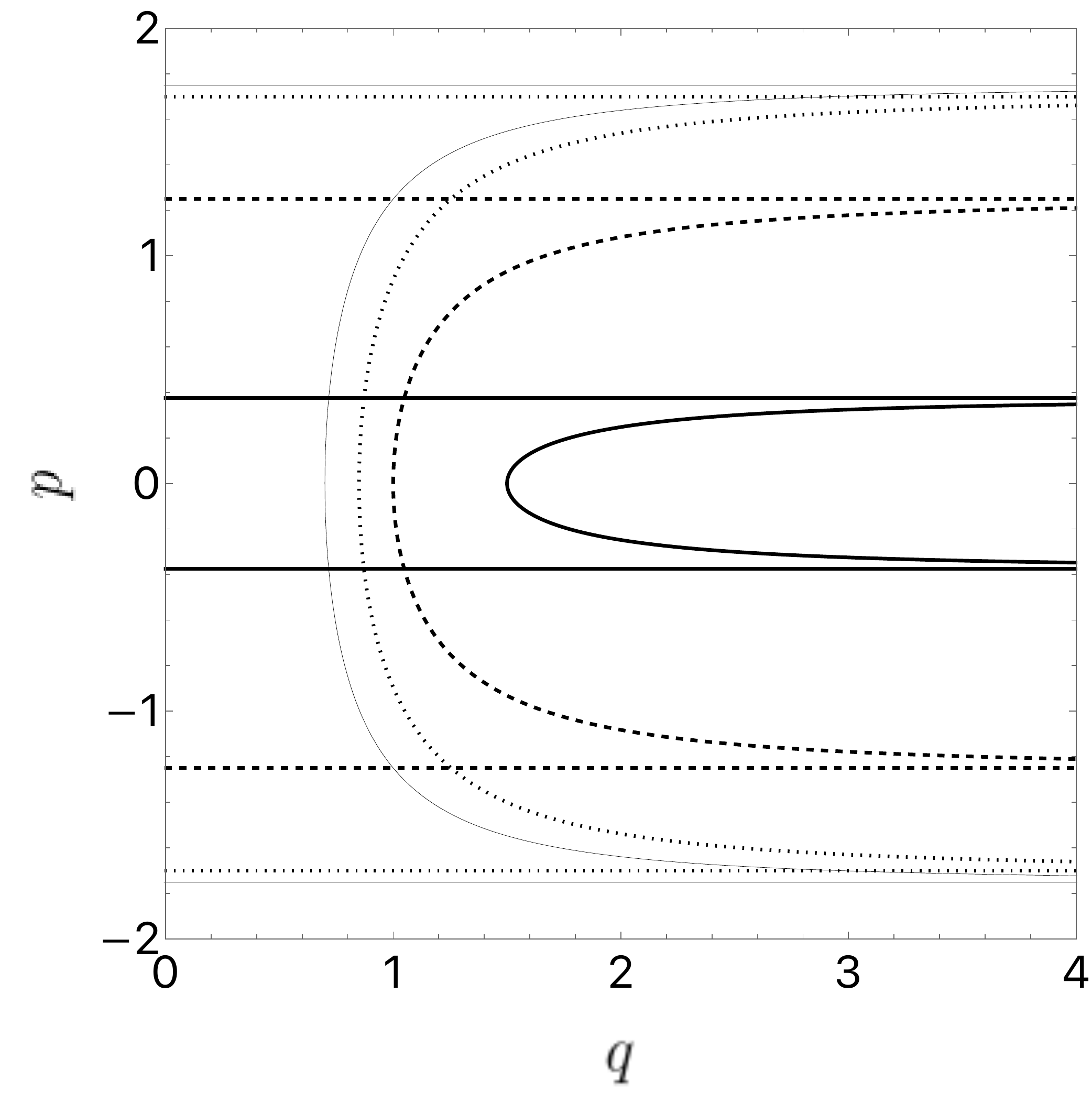}
 \caption{Background phase space evolutions: the straight lines
   represent Eqs.~\eqref{qpSol}, either going to or emerging from a
   singularity ($q\to 0$), while the curves are the solutions
   \eqref{solsem} leading to the same asymptotic classical lines. The
   semiclassical solution are seen to consist of a bounce smoothly
   joining expanding ($p>0$) and contracting ($p<0$) classical
   universes.}
\label{fig:Qtraj}
\end{figure}

It is very useful to have at disposal background solutions
$|\psi_\textsc{b}(\tau)\rangle$ corresponding to various energies and
with various spreads in $\widehat{Q}$ and $\widehat{P}$. One can find
a wide class of solutions by approximating the Hilbert space with a
family of coherent states built from a single wavefunction, the
so-called fiducial vector; this construction is presented in Appendix
\ref{A}.  For the present purpose, suffice it to note that any fixed
family of coherent states is given by state vectors $(q,p)\mapsto
|q,p\rangle$ in one-to-one correspondence with the phase space. In
practice, from a fiducial state $|\tilde\xi\rangle$, for which
$\langle\tilde\xi|\widehat{Q}|\tilde\xi\rangle =1$ (recall $q$,
and therefore $\widehat{Q}$, is dimensionless) and
$\langle\tilde\xi|\widehat{P}|\tilde\xi\rangle =0$, one builds
the coherent state through \cite{Klauder:2015ifa}
\begin{equation}
|q(\tau),p(\tau)\rangle =
\ex^{i p(\tau) \widehat{Q}/\hbar}\ex^{-i \ln q(\tau) \widehat{D}/\hbar}|\tilde\xi\rangle,
\label{CohSta}
\end{equation}
where
$\widehat{D}=\frac{1}{2}(\widehat{Q}\widehat{P}+\widehat{P}\widehat{Q})$
is the dilation operator. The expectation values of $\widehat{Q}$ and
$\widehat{P}$ in $|q(\tau),p(\tau)\rangle$ are respectively $q(\tau)$
and $p(\tau)$.

The dynamics confined to the vectors $|q(\tau),p(\tau)\rangle$ can be
deduced from the quantum action
\begin{equation}
\mathcal{S}_\textsc{b} =\int\langle q(\tau),p(\tau)|\left(i\hbar
\frac{\partial}{\partial \tau} - \widehat{H}^{(0)}\right)
|q(\tau),p(\tau)\rangle\dd \tau,
\end{equation}
which, upon using the properties of the state \eqref{CohSta}, can be
transformed into
\begin{equation}
\mathcal{S}_\textsc{b} =\int \left\{ \dot{q}(\tau) p(\tau) -
H_\text{sem}\left[q(\tau),p(\tau)\right] \right\}\dd \tau,
\end{equation}
with the semiclassical Hamiltonian given by
\begin{equation}
 H_\text{sem} =\langle q,p|\widehat{H}^{(0)}|q,p\rangle,
\end{equation}
from which one derives the ordinary Hamilton equations
\begin{equation}
  \dot{q}=\frac{\partial {H}_\text{sem}}{\partial p} \quad \text{and}
  \quad \dot{p}=-\frac{\partial {H}_\text{sem}}{\partial q}.
\label{eomsem}
\end{equation}
Given the quantum Hamiltonian \eqref{qh0}, we find that the
semiclassical background Hamiltonian reads, by virtue of our ordering
choice \eqref{ClassQsym} (with $\alpha=0$)
\begin{equation}
H_\text{sem}=2\kappa_0\left(p^2+\frac{\hbar^2\mathfrak{K}}{q^2}\right),
\label{HK}
\end{equation}
where the constant $\mathfrak{K}$ is positive
($\mathfrak{K}>0$), irrespective of whether ${\mathfrak{c}_{0}}=0$ or
${\mathfrak{c}_{0}}>0$.  The actual value of $\mathfrak{K}$ depends on
the choice of family of coherent states, as illustrated in Appendix
\ref{A}. We find the solution to \eqref{eomsem} to read
\begin{subequations}
\label{solsem}
\begin{align}
  &q=q_\textsc{b} \sqrt{1+(\omega\tau)^2}, \\ p&= \frac{q_\textsc{b}
    \omega^2}{4\kappa_0} \frac{\tau}{\sqrt{1+(\omega\tau)^2}},
\end{align}
\end{subequations}
where $q_\textsc{b}^2 = 2 \kappa_0 \hbar^2\mathfrak{K}/H_\text{sem}$ and
$\omega = 2 H_\text{sem}/(\hbar\sqrt{\mathfrak{K}})$. We display in
Fig. \ref{fig:Qtraj} a few trajectories in the phase space
illustrating these solutions.

\begin{figure}[t]
\centering
 \includegraphics[width=0.45\textwidth]{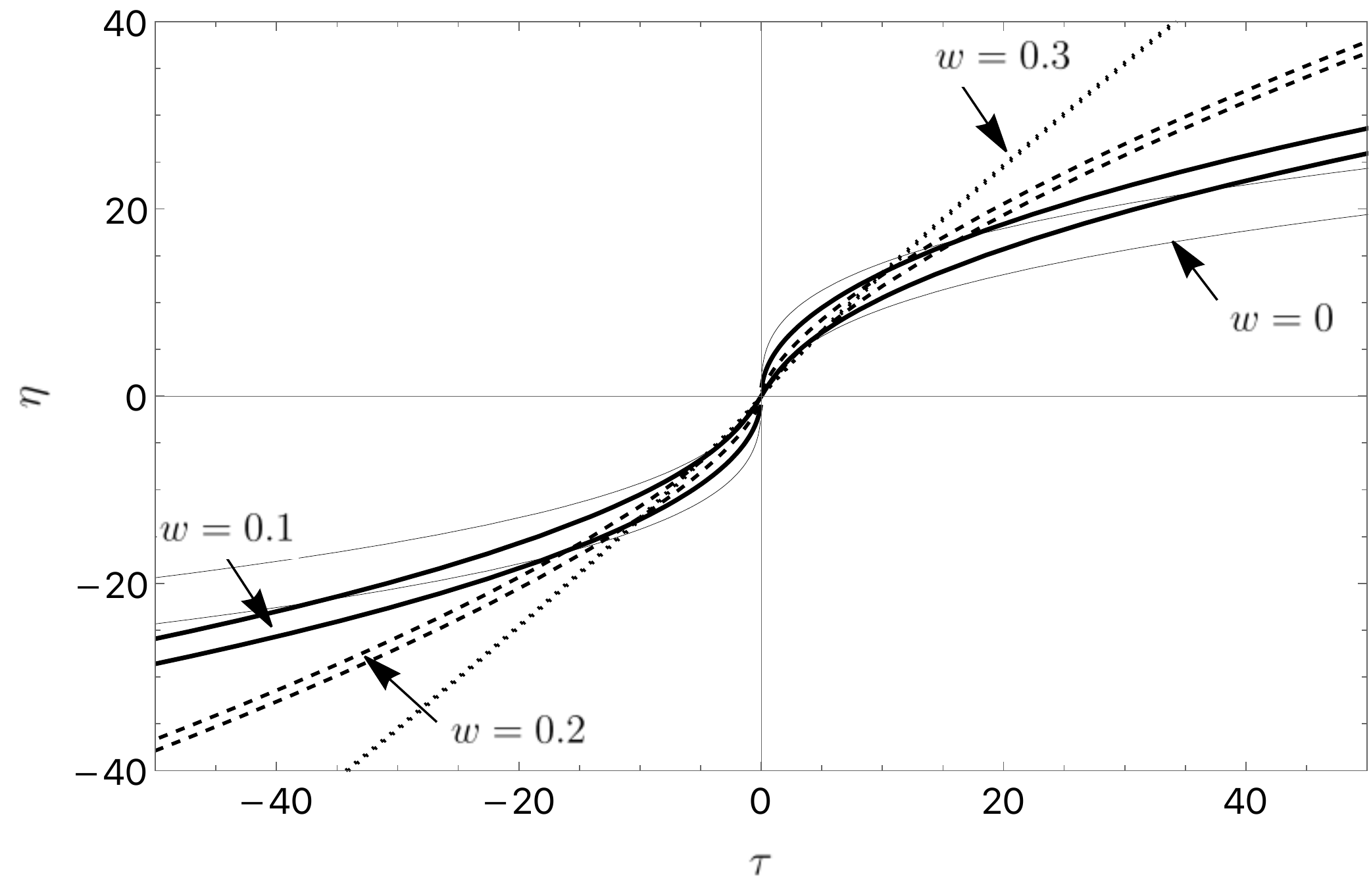}
\caption{Conformal time $\eta$ as a function of $\tau$, for the
  classical \eqref{etaCl} and quantum \eqref{etaSem} solutions for
  $w=0$ (thin line), $w=0.1$ (thick), $w=0.2$ (dashed) and $w=0.3$
  (dotted). The quantum conformal time tends to the classical one up
  to a constant factor, which vanishes for $=\frac13$.}
\label{figEta}
\end{figure}

With this semiclassical solution, one can also integrate \eqref{eta}
to yield the conformal time $\eta$, as a function of $\tau$
\begin{equation}
\eta = (1+w) \tau \left( \frac{q_\textsc{b}}{\gamma}\right)^{2 r_1} \,
\mathcal{F}\left[ \frac12,-r_1;\frac32;-\left(\omega \tau\right)^2 \right],
\label{etaSem}
\end{equation}
where $\mathcal{F}(a,b;c;z)$ is the hypergeometric function (see Sec.~15 of
Ref.~\cite{638211}).  As expected, one recovers the classical power
law \eqref{etaCl} in the large-time classical limit $\tau \gg
\omega^{-1}$, up to a constant depending on the equation of state $w$
and vanishing for $w=\frac13$. Fig.~\ref{figEta} shows the classical
and quantum relationships $\eta(\tau)$.

\subsection{Perturbation modes}

Given that the dynamics of the background state is fixed by
$|\psi_\textsc{b}\rangle \to |q(\tau),p(\tau)\rangle$, the dynamics of
the perturbation state $|\psi_\textsc{p}(\tau)\rangle$ can be deduced
from the quantum action at second order $\mathcal{S}^{(0)+(2)} =
\mathcal{S}_\textsc{b}+\mathcal{S}_\textsc{p}$
\begin{equation}
  \mathcal{S}^{(0)+(2)} =\int\langle \psi(\tau) |
  \left(i\hbar\frac{\partial}{\partial \tau} - \widehat{H}^{(0)}
  +\sum_{\bm{k}}\widehat{H}^{(2)}_{\bm{k}}\right)
  |\psi(\tau)\rangle\dd \tau,
\end{equation}
with the state vector given by \eqref{psiBP}. Extracting the zeroth
order action $\mathcal{S}_\textsc{b}$, one finds
\begin{equation}
  \mathcal{S}_\textsc{p} = \int \langle \psi_\textsc{p}|
  \left(i\hbar\frac{\partial}{\partial \tau} +\sum_{\bm{k}}
  \widehat{H}^{(2)}_{\bm{k}} \right) |\psi_\textsc{p} \rangle\dd \tau,
\end{equation}
and setting $|\psi_\textsc{p} \rangle = \prod_{\bm{k}}
|\psi_{\bm{k}}\rangle$ with
$\langle\psi_{\bm{k}_1}|\psi_{\bm{k}_2}\rangle =
\delta_{\bm{k}_1,\bm{k}_2}$, one gets the associated Schr\"odinger
equation for each Fourier mode $|\psi_{\bm{k}}\rangle$ (up to an
irrelevant phase factor), namely
\begin{equation}
  i\hbar\frac{\partial}{\partial \tau}|\psi_{\bm{k}}\rangle =
  \tilde{H}_{\bm{k}} |\psi_{\bm{k}}\rangle,
\end{equation}
where the operator $\tilde{H}_{\bm{k}} \equiv - \langle q,p|
\widehat{H}^{(2)}_{\bm{k}}|q,p\rangle$ is obtained from either
\eqref{qham} or \eqref{H2kQ} depending on the choice of
parametrization. In the former case, the second-order Hamiltonian
generating the dynamics of perturbations in the fluid
parametrization reads
\begin{equation}\label{HAMsem}
  \langle q,p|\widehat{H}^{(2)}_{\bm{k}}|q,p\rangle=
  \frac{1}{2}|\widehat{\pi}_{\phi,{\bm{k}}}|^2 +\frac{\Ls}{2}w(1+w)^2
  \left(\frac{{q}}{\gamma}\right)^{4r_1}
  k^2|\widehat{\phi}_{\bm{k}}|^2,
\end{equation}
where the value of $\Ls$ depends on the value of $\Lq$ and the family
of coherent states used to approximate the background dynamics.

The
Heisenberg equations of motion read
\begin{subequations}\begin{align}
    \frac{\dd}{\dd\tau}\widehat{\phi}_{\bm{k}}
    &=-\widehat{\pi}_{\phi,{\bm{k}}},\label{HeisPhi}\\
    \frac{\dd}{\dd\tau}\widehat{\pi}_{\phi,{\bm{k}}}&=\Ls
    w(1+w)^2
    \left(\frac{{q}}{\gamma}\right)^{4r_1}
    k^2\widehat{\phi}_{\bm{k}},
    \label{HeisPiPhi}
    \end{align}
\end{subequations}
and it follows from \eqref{HeisPhi} that
$$\widehat{\pi}_{\phi,\bm{k}} = \sqrt{\frac{\hbar}{2}}\left[
  a_{\bm{k}}\dot{\phi}^*_k(\tau)+ a^{\dagger}_{-{\bm{k}}}
  \dot{\phi}_k(\tau)\right],$$ and hence the canonical commutation
rule, namely
$[\widehat{\phi}_{-\bm{k}},\widehat{\pi}_{\phi,\bm{k}}]=i\hbar$,
implies the normalization condition on the mode functions
$\dot{\phi}_k \phi^*_{k}- \phi_k \dot{\phi}^*_k = 2 i$. By combining
the above equations, we may obtain the second-order dynamical equation
for $\widehat{\phi}_{\bm{k}}$, which must also be obeyed by the mode
function $\phi_k$. We switch to the internal conformal clock given by
Eq.~\eqref{eta} and rescale the mode functions, $v_k^\N = Z \phi_k$,
where $v_k^\N $ is the Mukhanov-Sasaki variable. The superscript
``\textsmaller{F}'' indicates that its dynamics is generated by the
fluid Hamiltonian. More specifically, we find that the dynamics of
$v_k^\N$ generated by the Hamiltonian \eqref{HAMsem} reads
\begin{equation}
  \frac{\dd^2 v_k^\N }{\dd\eta^2}+\left[ k_\N^2- \mathcal{V}_\N (\eta)
    \right] v_k^\N =0,
\label{nqeom}
\end{equation}
with the effective wave number $k_\N \equiv \sqrt{\Ls w} k$, and the
fluid potential given by
\begin{equation}
   \mathcal{V}_\N = \frac{8}{9q^2 Z^4}
   \frac{(2\kappa_0)^2(1-3w)}{(1-w)^2}
  \left[ p^2-\frac{{3 (1-w) \mathfrak{K}}}{2 q^2}
  \right].
\label{VN}
\end{equation}

\begin{figure}
\includegraphics[width=0.45\textwidth]{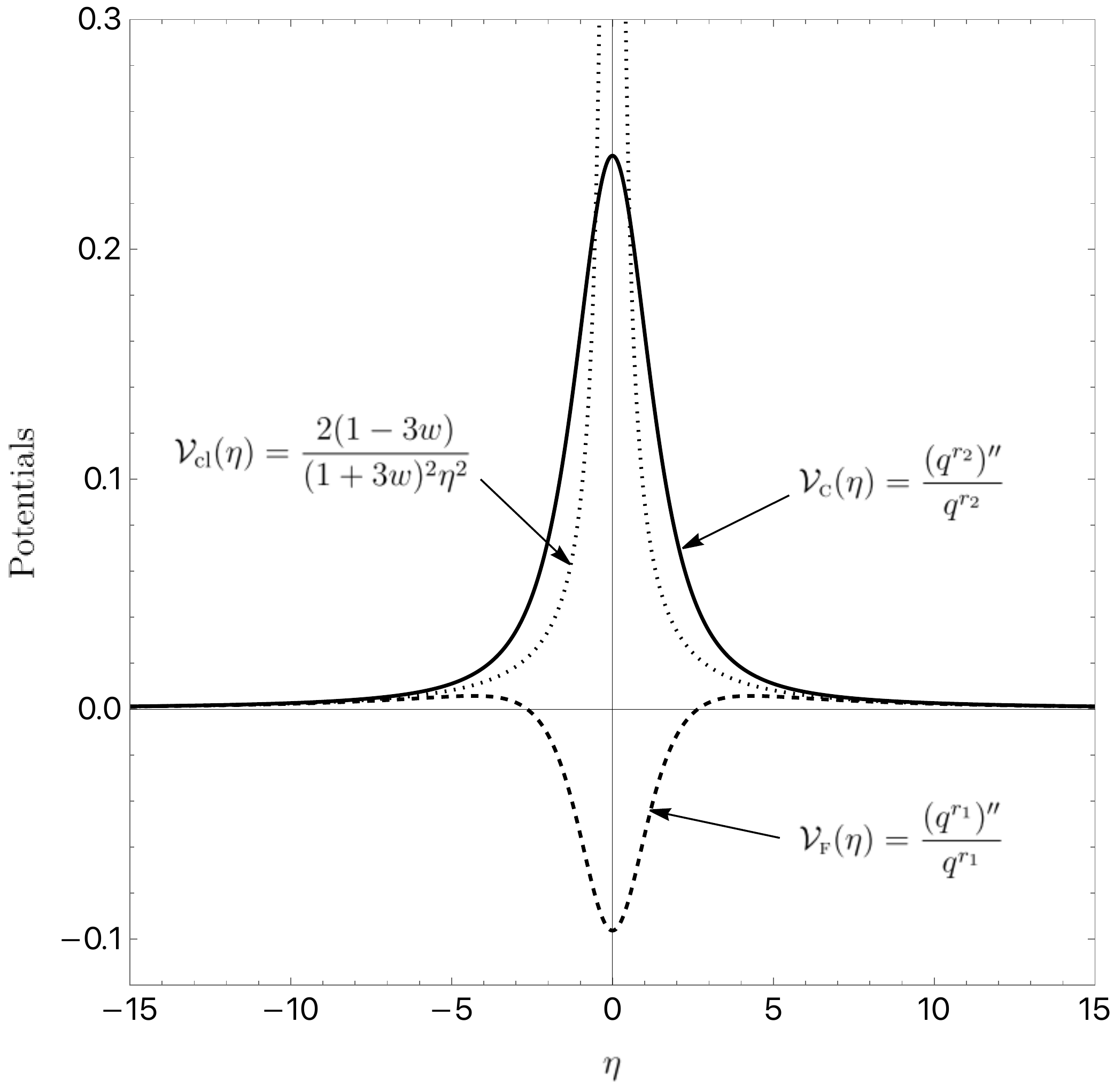}
\caption{The gravitational potentials $V_\MS$ (full line),
  from \eqref{msqeom}, and $V_\N$ (dashed line), from
  \eqref{nqeom}, as functions of the conformal time $\eta$; the
  parameter values are chosen as $q_\textsc{b} \to 1$, $\omega \to 1$,
  $\kappa_0 \to 1$ and $w=0.2$ for the purpose of illustration. These
  potentials are deduced from the quantum fluid \eqref{HAMsem} and
  conformal \eqref{MSqh} Hamiltonians and the classical
  Hamiltonian. They all asymptotically decay as $\eta^{-2}$ far from
  the bounce where they are well-approximated by their classical
  counterpart given by $V_\text{cl} = \frac{2(1-3w)}{(1+3w)^2\eta^2}$
  (dotted line) [cf. Eq.~\eqref{Vcl}].}
\label{three}
\end{figure}

Note that for large $q$, i.e. away from the bounce, the semiclassical
correction becomes negligible so that the semiclassical potential
\eqref{nqeom} approaches the classical one \eqref{HamMS}. Indeed,
using $\dot{Z}/Z = r_1 \dot{q}/q$ and $q' = \dot{q}/Z^2$, one finds
$$
\frac{Z''}{Z} = \frac{r_1}{Z^4} \left[ \frac{\ddot{q}}{q} -
(1+r_1) \left(\frac{\dot{q}}{q}\right)^2\right],
$$
and replacing the function $q(\tau)$ by the solution \eqref{solsem}
for the background semiclassical trajectory, it is straightforward to
check that, for all times, the potential $\mathcal{V}_\N$ can be given
the familiar form $\mathcal{V}_\N = Z''/Z = \left( q^{r_1}\right)'' /
q^{r_1}$. Since the semiclassical trajectory \eqref{solsem} is
asymptotic to the classical one \eqref{qpSol} for
$\omega\tau\to\infty$, i.e.  for $\eta\to\infty$, the fluid
potential satisfies
$$
\lim_{\eta\to \infty} \mathcal{V}_\N (\eta) = \mathcal{V}_\text{cl}
(\eta),
$$
where $\mathcal{V}_\text{cl}$ is given by \eqref{Vcleta}; it is
illustrated in Fig.~\ref{three}.

\begin{figure}
\includegraphics[width=0.45\textwidth]{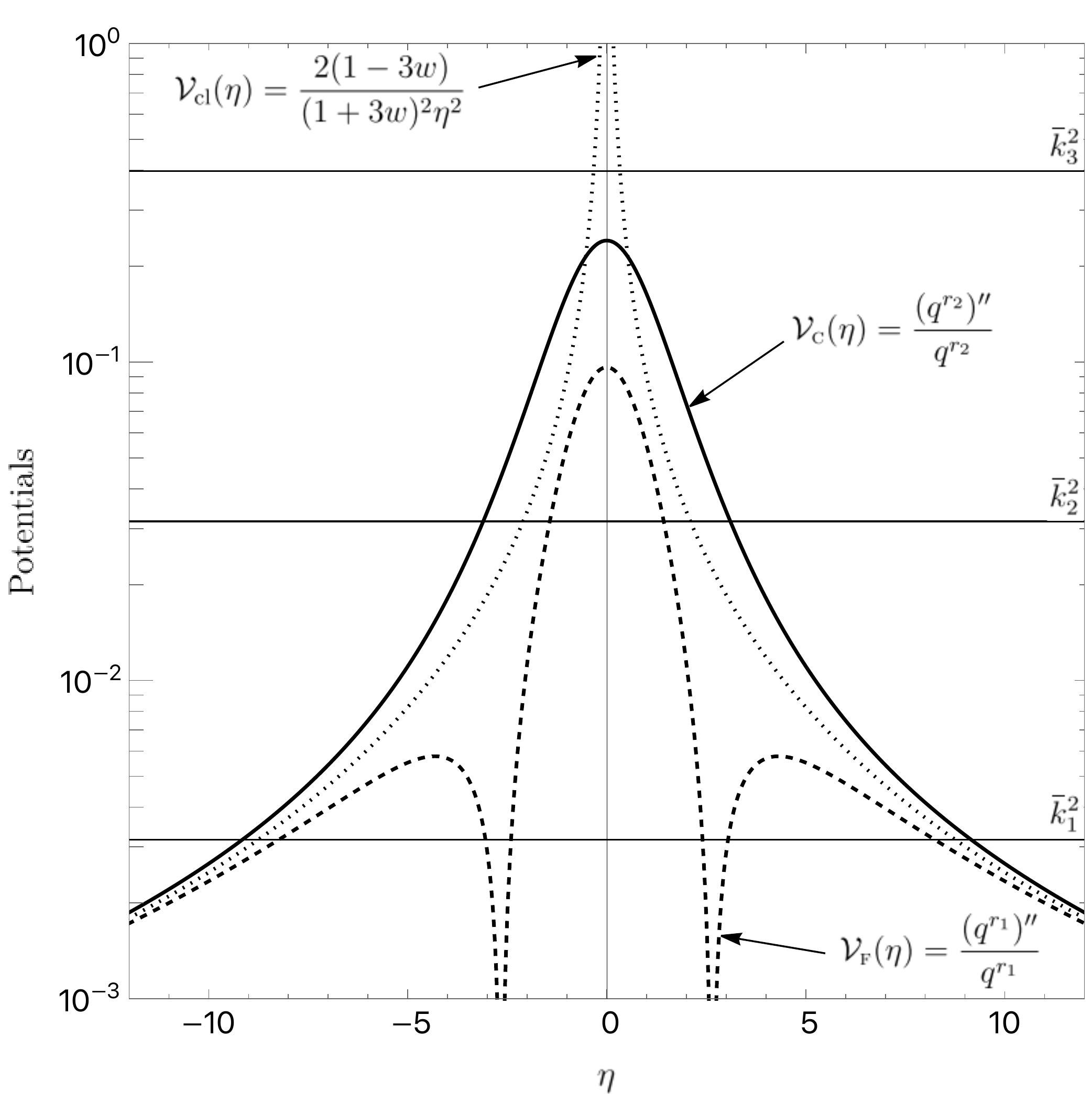}
\caption{Same as Fig.~\ref{three} in logarithmic scale for the
  potentials, with different wave numbers ($\bar{k}$ standing for
  either $k_\N$ or $k_\MS$ depending on the case at hand),
  illustrating the various possible predictions.  For $\bar{k} \sim
  \bar{k}_3$, the quantum potentials is not felt by the perturbations,
  and only the classical potential induce a non trivial spectrum.  In
  the region of wavelengths around $\bar{k}\sim \bar{k}_2$, the
  perturbations enter the potentials at different points, but the
  characteristic behaviour is more or less comparable; one would
  expect in this regime to have different amplitudes and even perhaps
  power indices, but an overall similar shape. For $\bar{k}\sim
  \bar{k}_1$ on the other hand, the number of entries and exits of the
  perturbation in and out of the potentials $V_\N$ and $V_\MS$ being
  different, predictions between the two models could radically
  differ, e.g. with superimposed oscillations changing the shape of
  the primordial power spectrum.}  
\label{threeBis}
\end{figure} 

The same procedure applied to the conformal parametrization
yields
\begin{equation}
  \langle q,p|\widehat{H}^{(2)}_{\bm{k}}|q,p\rangle=
  \frac12 Z^2 \Ms\, \left( |\widehat{\pi}_{v,k}|^2 +
  \Omega^2_v |\widehat{v}_k|^2 \right),
\label{MSqh}
\end{equation}
with
\begin{equation}
  \Omega^2_v = wk^2-\frac{8\Ms^{-1}}{9q^2 Z^4}
  \frac{(2\kappa_0)^2(1-3w)}{(1-w)^2}
  \left( \Ns p^2+
  \frac{ \hbar^2\Ts}{q^2}
  \right),
\label{MSqhHsem}
\end{equation}
where $\Ms$, $\Ns$, $\Ts$ depend on the family of coherent states used
to approximate the background dynamics and on the values of $\Mq$,
$\Nq$, $\Rq$ and $\Tq$ and $\Nq$, respectively.  For the following discussion, one
should bear in mind that all the quantities $\Ks$, $\Ls$, $\Ms$, $\Ns$
and $\Ts$ are positive definite. The canonical commutation rule
implies the normalization condition on the mode functions
$$\dot{v}_k v^*_k-v_k \dot{v}^*_k=2i(1+w)
\left(\frac{q}{\gamma}\right)^{-2r_1}
\Ms = 2 i Z^2 \Ms.$$
After switching to the internal conformal clock, the normalization
condition reads
$v'_k v^*_k-v_k v^{*\prime}_k=2i\Ms$
and the Hamiltonian \eqref{MSqh} is found to generate the
following dynamics of the mode function $v_k^\MS $ (the
subscript ``\MS'' now indicating that its
dynamics is generated by
the conformal Hamiltonian)
\begin{equation}
  \frac{\dd^2 v_k^\MS }{\dd\eta^2}+\left[  \Ms^2 w k^2
  - \Ms \Ns \mathcal{V}_\MS (\eta) \right] v_k^\MS =0,
\label{msqeom}
\end{equation}
where the potential reads
\begin{equation}
   \mathcal{V}_\MS = 
   \frac{8}{9q^2 Z^4}
  \frac{(2\kappa_0)^2(1-3w)}{(1-w)^2}
  \left( p^2+ \displaystyle\frac{\hbar^2\Ts/\Ns}{q^2}
  \right),
\label{VMS}
\end{equation}
whose limit for large $q$ yields back
the classical case \eqref{Vclass}. The usual
Mukhanov-Sasaki equation is recovered from \eqref{msqeom}
provided one defines a rescaled conformal time $\varsigma$
through $\varsigma = \sqrt{\Ms \Ns} \eta$,
leading to
\begin{equation}
  \frac{\dd^2 v_k^\MS }{\dd\varsigma^2}+\left[  k_\MS^2
  - \mathcal{V}_\MS (\varsigma) \right] v_k^\MS =0,
\label{msqvarsigma}
\end{equation}
as expected; in Eq.~\eqref{msqvarsigma}, the effective wave number is
$k_\MS \equiv k \sqrt{w \Ms/\Ns}$.

\begin{figure}
\includegraphics[width=0.45\textwidth]{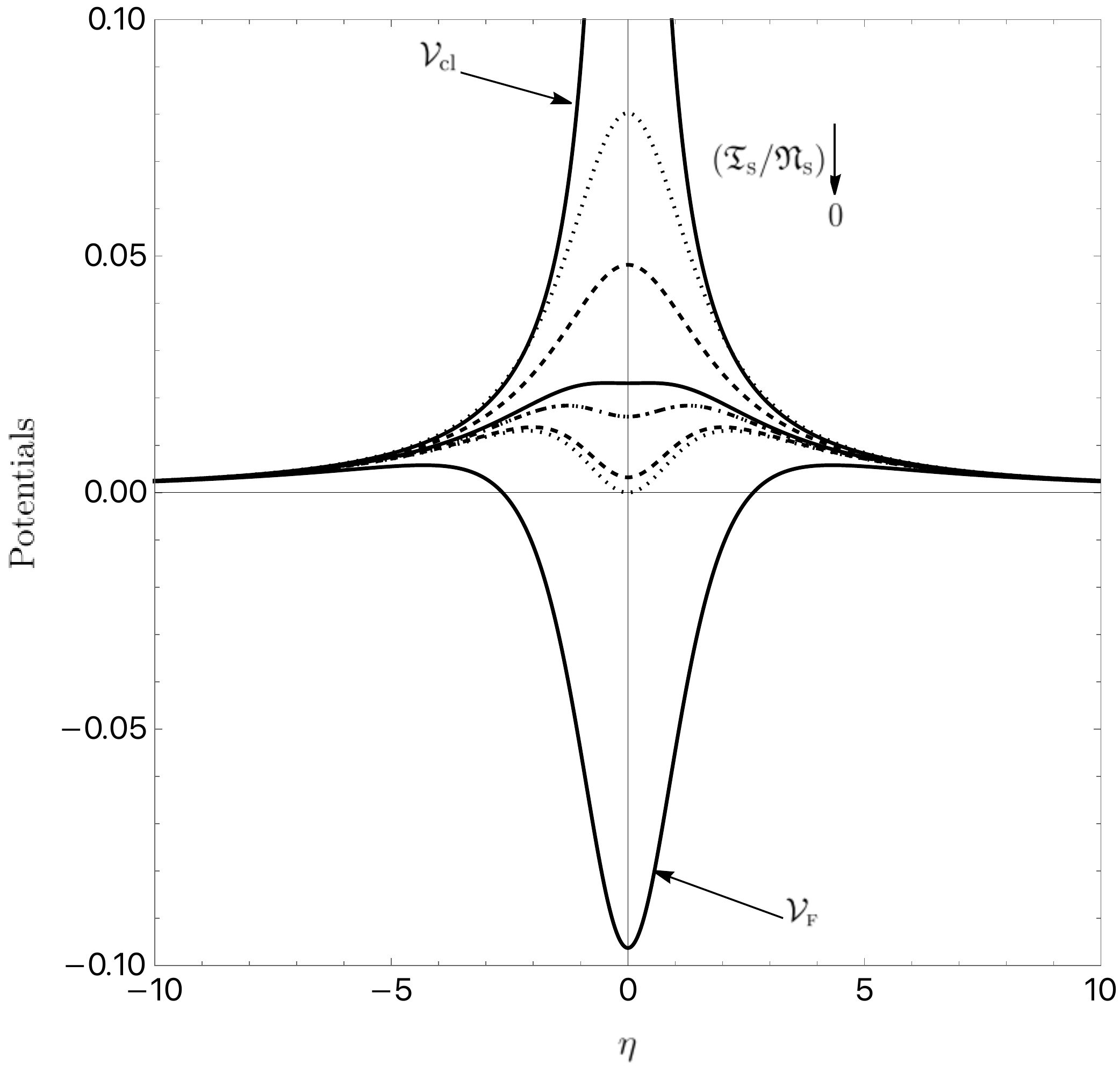}
\caption{Shape of the conformal potential $\mathcal{V}_\MS$ 
\eqref{VMS} for $w=0.2$ and various values of $\Ts/\Ns$, decreasing
to 0 according to the arrow, compared with the classical
$\mathcal{V}_\text{cl}$ \eqref{Vcleta} and fluid
$\mathcal{V}_\N$ \eqref{VN} potentials. The special value
(we assume $ \mathfrak{K}\to 1$)
$\Ts/\Ns = 3 (1-w)/(1-3w) \simeq 0.375$ (not shown) corresponds to that in 
Fig.~\ref{three}
for which $\mathcal{V}_\MS = \left( q^{r_2} \right)''/q^{r_2}$. Shown
are $\Ts/\Ns = 0.125$ (dotted line), $0.075$ (dashed), $0.036$ (full),
$0.025$ (dot-dashed), $0.005$ (dashed) and $0$ (dotted). The
full line represents a critical point above which the potential
has only one maximum. For $\Ts/\Ns=0$, the potential is minimum
at the bounce where it vanishes.}
\label{fig_Shapes}
\end{figure} 

We have seen above that $\mathcal{V}_\N = \left( q^{r_1}\right)'' /
q^{r_1}$. Let us see under what conditions the potential
$\mathcal{V}_\MS$ can also be put in the form $X''/X = \left(
q^r\right)'' / q^r$ for a given function $X(\eta) = q^r$ with a power
$r$ to be determined. Straightforward calculation yields
\begin{equation}
\begin{aligned}
\frac{X''}{X} & = \frac{r}{Z^4} \left[ \frac{\ddot{q}}{q}
+ (r-2r_1-1) \left( \frac{\dot{q}}{q} \right)^2
\right] \nonumber \\
& = \frac{4 (2\kappa_0)^2}{Z^4 q^2} r (r-2 r_1 -1) \left[
p^2 + \frac{\mathfrak{K}}{(r-2r_1-1) q^2}
\right], \nonumber
\end{aligned}
\end{equation}
where in the second equality we have made use of the semiclassical
solution \eqref{solsem}. In order to recover the classical limit
\eqref{vVcl}, the power $r$ should satisfy $r(r-2r_1+1) = \frac29
(1-3w)/(1-w)^2$, whose two roots happen to coincide with $r_1$ and
$r_2$. Setting $r=r_1$ yields \eqref{VN}, with a negative coefficient
in the $q^{-2}$ term (we assume $0<w<1$), as could have been
anticipated. The second root $r=r_2$ yields instead a positive
coefficient in the $q^{-2}$ term, and reproduces \eqref{VMS} if we
demand that $w<\frac13$ and 
\begin{equation}
\frac{\Ts}{\Ns} = \frac{3\mathfrak{K}(1-w)}{1-3w}\quad
\Longrightarrow \quad
\mathcal{V}_\MS \to \frac{(q^{r_2})''}{q^{r_2}}.
\label{TsNs}
\end{equation}
Both potentials are shown in Figs.~\ref{three} and \ref{threeBis}.

It is clear from \eqref{VN} and \eqref{VMS} that the two equivalent
parametrizations of the classical model induce two inequivalent
quantum theories, as is clear from Figs.~\ref{three} and
\ref{threeBis} showing a comparison of the respective gravitational
potentials. The difference is perhaps even clearer when the
gravitational potentials are given in the familiar form based in the
configuration space and the semiclassical variable $q$ is raised to
two distinct powers, i.e. $r_1=\frac{3w-1}{3(1-w)}$ and $r_2 =
\frac{2}{3(1-w)}$.  In some sense these two parametrizations are
exhaustive in regard to the quantization ambiguity as these are the
only powers possible for theories that satisfy the classical limit, as
follows from our discussion below \eqref{eomP}. The source of the
ambiguity is the non-linearity of the theory of gravity. Since the
quantization concerns both the linear perturbations and the background
variables, the transformation of the perturbation
variables \eqref{ntoms} is nonlinear, contrary to the situation of 
Ref.~\cite{Martin:2007bw}, and therefore, it leads to
unitarily inequivalent theories.

\section{Conclusion}

We have suggested a finite cosmological model in which quantum
gravitational effects play a leading role, resolving the classically
expected singularity to a bouncing scenario. Our model consists in
adding to general relativity a perfect fluid with constant equation
of state $w$. Classically, the FLRW solution initiates out of or
contracts to a singularity at which the scale factor $a$ vanishes. The
perturbations around such a background also tend to diverge at the
singularity.

Upon quantizing the background, factor ordering ambiguities permit
to add to the zeroth order Hamiltonian a repulsive potential term.
Assuming a coherent state to describe the semiclassical evolution,
one can then calculate a phase space trajectory which, thanks to
the quantum effective potential, smoothly connects the contracting
and expanding solutions, avoiding the singularity in the process.

Most models then would identify these bouncing trajectories as
classical, and would then go on to quantize the perturbations
on top. By doing so, one would then be allowed whatever canonical
transformation on the perturbation variables, leading to
classically undistinguishable theories. Here however, we take
seriously the quantum nature of the background time development
and show that the classically harmless canonical transformations
become unitarily inequivalent theories with potentially different
physical predictions.

\begin{acknowledgments}
P.M. and J.C.M. acknowledge the support of the National Science Centre
(NCN, Poland) under the research grant 2018/30/E/ST2/00370. The project
is cofinanced by the Polish National Agency for Academic Exchange and
PHC POLONIUM 2019 (Grant No. 42657QJ). We thank Claus Kiefer and
J\'er\^ome Martin for illuminating discussions.
\end{acknowledgments}

\appendix

\section{Affine coherent states and affine quantization}\label{A}

In what follows we discuss the affine coherent states and their
application to affine quantization and to semiclassical description of
dynamics
\cite{Fresneda:2015fla,Gazeau:2009zz,Bergeron:2013ika,Bergeron:2017ddo}.
% For simplicity, we set $\hbar=1$ in the appendices.

\subsection*{Coherent states and quantization}

The background phase space $(q,p)$ is the half-plane that is not
invariant under the usual group of $q-$ and $p-$translations. For this
reason the application of ``canonical quantization" based on the
unitary and irreducible representation of the group of translations,
the Weyl-Heisenberg group, is problematic. It is however possible to
consider a more general quantization that is based on any minimal
group of canonical transformations that enjoys a nontrivial unitary
representation, the so-called covariant integral quantization. In the
case of the half-plane the natural choice is the affine group of a
real line, $(q,p)\in\mathbb{R}^+\times\mathbb{R}$,
\begin{equation}
\left( q',p'\right) \circ \left( q,p\right) = 
\left( q'q,\frac{p}{q'}+p'\right).
\end{equation}
Its unitary, irreducible and square-integrable representation in the
Hilbert space $\mathcal{H}=L^2(\mathbb{R}^+,\dd x)$ reads
\begin{equation}
\langle x|U(q,p)|\psi\rangle = \langle x|q,p\rangle
=\frac{\ex^{ipx/\hbar}}{\sqrt{q}}\psi\left(
\frac{x}{q}\right),
\label{qpx}
\end{equation}
where $\psi(x)=\langle x|\psi\rangle\in \mathcal{H}$. 

Let us consider a particular example of the covariant integral
quantization that is based on coherent states. In the present case,
they are the affine coherent states,
\begin{equation}\label{affcs}
\mathbb{R}^+\times\mathbb{R}\ni (q,p)\mapsto |q,p\rangle:=U(q,p)|\xi\rangle
\in\mathcal{H},
\end{equation} 
where $|\xi\rangle$ is the so-called fiducial vector, a fixed
normalized vector in Hilbert space such that $\mathcal{N} = \rho(0)
<\infty$, with
$$
\rho(\alpha) = \int \frac{|\xi(x)|^2}{x^{\alpha+1}}\, \dd x,
$$
and the operator $U(q,p)$ is given by Eq.~\eqref{CohSta}. The resolution of
unity is
\begin{align}
\int \frac{\ud q \ud p}{2\pi \hbar \mathcal{N}} | q,p\rangle\langle q,p| = \mathbb{1},
\end{align}
as can be verified in a straightforward manner using Eq.~\eqref{qpx}
and applying the above operator on two arbitrary states $\langle \phi_1|$ and
$|\phi_2\rangle$:
$$
\int \frac{\ud q \ud p}{2\pi\hbar \mathcal{N}} \langle \phi_1| q,p\rangle\langle 
q,p|\phi_2 \rangle =
\int \dd x \phi_1^* (x) \phi_2(x) = \langle \phi_1|\phi_2\rangle,
$$
using the usual closure relation
$$
\int \dd x |x\rangle\langle x| = \mathbb{1}
$$
and the property
$$
\delta (x-y) = \int \frac{\dd p}{2 \pi\hbar} \ex^{i p (x-y)/\hbar}
$$
for the Dirac distribution.

The affine coherent state quantization is obtained by substituting functions of
$q$ and $p$ are replaced by
\begin{equation}\label{qmap}
  f(q,p)\mapsto A_f:=\int_{\mathbb{R}^+\times\mathbb{R}} \frac{\dd
    q\dd p}{2\pi \hbar\mathcal{N}}|q,p\rangle\, f(q,p)\,\langle q,p|,
\end{equation}
with $\mathcal{N}$ the normalization constant. Let us also introduce
\begin{equation}
\sigma(\alpha)=\int \left| \frac{\dd \xi(x)}{\dd x}
\right|^2 \frac{\dd x}{x^{\alpha+1}},
\end{equation}
which is the same as $\rho$ with the function $\xi(x)$ replaced by its
derivative $\xi'(x)$.

One may easily find the affine coherent state quantization
(\ref{qmap}) of the following observables through (see, e.g.,
Appendices of \cite{Bergeron:2017ddo} or \cite{Almeida:2018xvj} 
for explicit computations)
\begin{subequations}
\label{Arules}
\begin{align}
A_1 & = \mathbb{1}, \label{AId}\\
A_{q^\alpha} & =\mathfrak{a}(\alpha) \widehat{Q}^\alpha, \label{Aqalpha}\\
A_p& = \widehat{P},\label{Ap}\\
A_{q^{\alpha}p^2}& = \mathfrak{a}(\alpha) \widehat{Q}^\alpha
    \widehat{P}^2 - i \alpha \hbar \mathfrak{a}(\alpha) \widehat{Q}^{\alpha-1}
    \widehat{P} + \mathfrak{c}(\alpha) \hbar^2\widehat{Q}^{\alpha-2},\label{Aqalphap2}
\end{align}
\end{subequations}
where $\widehat{Q}$ and $\widehat{P}$ are the `position' and
`momentum' operators on the half-line: Eqs.~\eqref{Aqalpha} and \eqref{Ap}
are to be understood as $\langle x | A_{q^\alpha} |\phi\rangle
=\mathfrak{a}(\alpha)
x^\alpha \phi(x)$ and $\langle x | A_p |\phi\rangle = - i \hbar \dd\phi/\dd x$,
where $\phi(x) := \langle x|\phi\rangle$.

The parameters $$\mathfrak{a}(\alpha) =
\frac{\rho(\alpha)}{\rho(0)}$$
and $$\mathfrak{c}(\alpha) = \frac12\alpha(1-\alpha) \mathfrak{a}(\alpha) +
\frac{\sigma(\alpha-2)}{\rho(0)}$$ 
are calculable for any real fiducial vector $\xi (x)$, which
should be chosen such that $\mathfrak{a}(1) = 1$, i.e. $\rho(1)=\rho(0)$
in order to ensure that $A_q = \widehat{Q}$ so that Eqs.~\eqref{Aqalpha}
and \eqref{Ap} implement the required
usual commutation relation $[A_q,A_p] =
[\widehat{Q},\widehat{P}] = i\hbar$.

From the above, it follows that
the application of the affine quantization \eqref{qmap} to the
background Hamiltonian \eqref{h0} yields
\begin{equation}\label{qh00}
H^{(0)}\mapsto \widehat{H}^{(0)}=2\kappa\left(\widehat{P}^2+
\hbar^2\mathfrak{c}_{0}\widehat{Q}^{-2}\right),
\end{equation}
with $\mathfrak{c}_{0} = \mathfrak{c}(0)= \sigma(-2)/\rho(0)$.

Furthermore, using again \eqref{r1r2}, one may easily calculate the
various constants appearing in the perturbation Hamiltonians,
namely 
\begin{equation}
\Lq=\frac{\rho(4 r_1)}{\rho(0)}
\end{equation}
for \eqref{qham}, as well as
\begin{equation}
\begin{aligned}
\Mq&=\frac{\rho(2 r_1)}{\rho(0)},\\
\Nq &=\frac{\rho(-2 r_2)}{\rho(0)}, \\
\Rq &=2 \hbar r_2\Nq, \\
\end{aligned}
\end{equation}
and
\begin{equation}
\Tq=-r_2(1+2r_2) \Nq + \frac{\sigma\left( -2r_2-2 \right)}{\rho(0)},
\end{equation}
which appear in \eqref{H2kQ}. Obviously, these parameters are to a
large extent free as the affine quantization depends on the fiducial
vector $|\xi\rangle$.  One might think about the coherent state
quantization based on the fiducial vector as a convenient method for
parameterizing natural ordering ambiguities.
 
\subsection*{Semiclassical approximation}

The most important application of the affine coherent states in the
present work is to derive a useful semiclassical description. As
discussed around Eq.~\eqref{CohSta}, one needs to ensure the so-called
physical centering condition $\langle \widehat{Q} \rangle = 1$, where
the expectation value is taken in the fiducial state. This condition
may not be satisfied by the state $|\xi\rangle$, already normalised
to enforce the canonical commutation relation, and so we
introduce a new real fiducial vector $|\tilde{\xi}\rangle$ and the
associated moments $\tilde{\rho}(\alpha) = \int_{\mathbb{R}^+}
\frac{\dd x}{x^{\alpha+1}}|\tilde{\xi}|^2$ and $\tilde{\sigma}(\alpha)
= \int_{\mathbb{R}^+} \frac{\dd x}{x^{\alpha+1}}|\tilde{\xi}'|^2$. We find
\begin{subequations}
\label{Srules}
\begin{align}\begin{split}
\langle q,p|\widehat{Q}^\alpha \widehat{P}^2|q,p\rangle & = \tilde{\rho}(-\alpha-1)q^{\alpha}p^2+i\alpha\tilde{\rho}(-\alpha)q^{\alpha-1}p\\
+&\left[\tilde{\sigma}(-\alpha-1) +\frac{\alpha(1-\alpha)}{2}\tilde{\rho}(-\alpha+1)\right]q^{\alpha-2}, \label{SQP2}\end{split}\\
\langle q,p|\widehat{Q}^\alpha \widehat{P}|q,p\rangle & =\tilde{\rho}(-\alpha-1)q^{\alpha}p+i\frac{\alpha}{2}\tilde{\rho}(-\alpha)q^{\alpha-1}, \label{SQP}\\
\langle q,p|\widehat{Q}^\alpha |q,p\rangle & = \tilde{\rho}(-\alpha-1)q^{\alpha}.\label{SQ}
\end{align}
\end{subequations}
Note that the special case $\alpha = 0$ in \eqref{SQ}
yields the normalisation $\langle q,p|q,p\rangle = \tilde{\rho}(-1)
= \langle \xi | \xi \rangle = 1$.

For the quantum Hamiltonian \eqref{qh00}, we introduce the following
semiclassical Hamiltonian
\begin{equation}
H_\text{sem} := \langle q,p|\widehat{H}^{(0)}|q,p\rangle
= 2\kappa_0\left(p^2+\frac{\hbar^2\mathfrak{K} }{q^2}\right),
\label{HsemK}
\end{equation}
where the new constant $\mathfrak{K}$ is given by
$\mathfrak{K}=\mathfrak{c}_0 \tilde{\rho}(1)+
\tilde{\sigma}(-2)$. As for perturbations, it is straightforward
to determine the constant 
in \eqref{HAMsem}, namely
\begin{equation}
\Ls= \Lq\tilde{\rho}\left( -4r_1-1\right),
\end{equation}
whereas one gets
\begin{equation}
\begin{split}
\Ms & = \Mq\tilde{\rho}\left( -2r_1-1\right),\\ 
\Ns & = \Nq\tilde{\rho}\left( 2r_2-1\right),
\end{split}
\end{equation}
and
\begin{equation}
\begin{split}
\Ts =  \Nq\tilde{\sigma}(2r_2-1) + \Tq\tilde{\rho}\left( 2r_2+1\right)
\end{split}
\end{equation}
for those appearing in \eqref{MSqh}.

\subsection*{Fiducial vectors}

For the sake of concreteness in the present discussion let us consider
some examples of fiducial vectors and the specific values of
$\rho(\alpha)$, $\sigma(\alpha)$, $\tilde{\rho}(\alpha)$ and
$\tilde{\sigma}(\alpha)$ that they produce.  We use two distinct
families of fiducial vectors for quantization and for semiclassical
approximation.  This is due to the fact that they satisfy special and
distinct conditions. Namely, the fiducial vectors for quantization are
such as to preserve the canonical commutation rule (on the half-line),
whereas the fiducial vectors for semiclassical approximations are such
as to yield the expectation values for the momentum and position
operators in any coherent state, aligned with the phase space point to
which a given coherent state is assigned.

We consider the following family of fiducial vectors for quantization
\begin{align}\label{fid1}
\xi_{\nu}(x)=\left(\frac{\nu}{\pi}\right)^{\frac{1}{4}}
\frac{1}{\sqrt{x}}\exp\left[-\frac{\nu}{2}
\left(\ln x-\frac{3}{4\nu}\right)^2\right],
\end{align}
where $\nu>0$ is assumed, and for which we obtain
the corresponding coefficients
\begin{align}\begin{split}
\rho_\nu (\alpha) & =\exp\left[ \frac{(\alpha-2)(\alpha+1)}{4\nu}
  \right],\\ \sigma_\nu (\alpha) &
=\left[\frac{\nu}{2}+\left(\frac{\alpha+2}{2}\right)^2\right]
\exp\left[ {\frac{\alpha(\alpha+3)}{4\nu}}\right],\end{split}
\end{align}
which are positive definite. As expected, one verifies that
$\rho_\nu(1) = \rho_\nu(0) = \ex^{-1/(2\nu)}$, as needed to ensure
the correct commutation relation between the position variable
and its associated canonical momentum. We also note that
$\langle \xi |\widehat{Q}|\xi\rangle = \rho_\nu(-2) = \ex^{3/(2\nu)}
\not= 1$, so the physical centering condition is not fullfilled by
this fiducial state.

\begin{figure}
\includegraphics[width=0.45\textwidth]{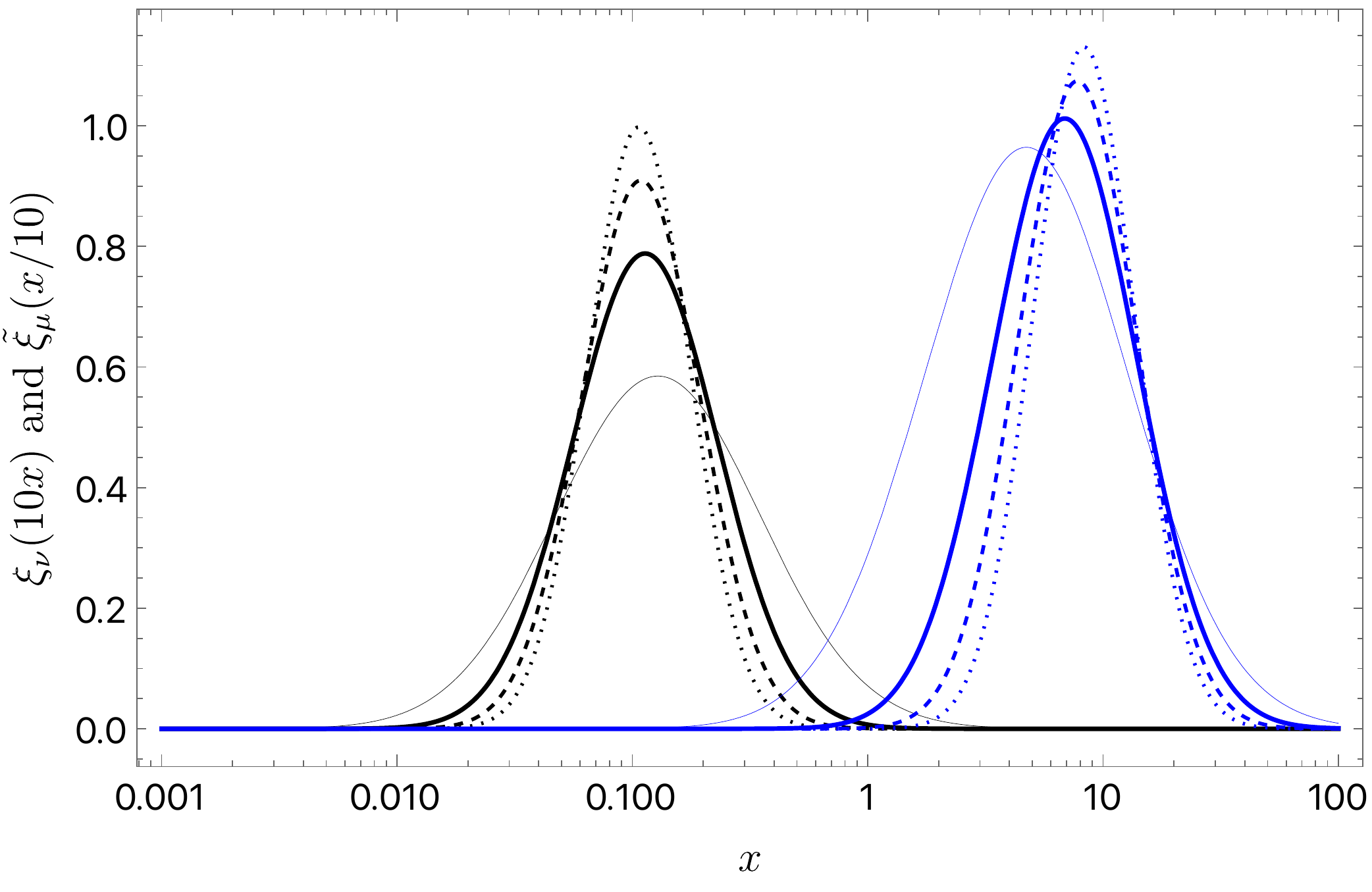}
\caption{Fiducial functions $\xi_\nu(10 x)$ and $\tilde\xi_\mu(x/10)$
(blue), for $\nu,\mu=1$ (thin line), $2$ (full), $3$ (dashed) and
$4$ (dotted). For better readability of the figure, the functions have
been shifted so that $\xi_\nu$
appears centered around $0.1$ and $\tilde\xi_\mu$ around $10$.
As functions of $x$, they should all be centered around $x=1$.}
\label{fig_Xis}
\end{figure} 

As for the semiclassical description, we consider the following family
of fiducial vectors
\begin{align}\label{fid2}
\tilde{\xi}_{\mu}(x)=\left(\frac{\mu}{\pi}\right)^{\frac{1}{4}}
\frac{1}{\sqrt{x}}\exp\left[-\frac{\mu}{2}\left(\ln
  x+\frac{1}{4\mu}\right)^2\right],
\end{align}
where now $\mu>0$ is assumed. In this case, we obtain
\begin{align}\begin{split}
\tilde{\rho}_\mu(\alpha)&=\exp\left[ \frac{(\alpha+1)(\alpha+2)}{4\mu}
  \right],\\ \tilde{\sigma}_\mu(\alpha)&=\left[\frac{\mu}{2}+
  \left(\frac{\alpha+2}{2}\right)^2\right]
\exp\left[ {\frac{(\alpha+3)(\alpha+4)}{4\mu}}\right].\end{split}
\end{align}
These are also positive definite as expected. It is now clear that
$\tilde\rho_\mu (-2) = 1$, as expected for the semiclassical
description, but that now $\tilde\rho (1)
= \ex^{3/(2\mu)} \not= \ex^{1/(2\mu)} = \tilde\rho(0)$ so that
these fiducial vectors cannot be used for quantization.

Some example functions
$\xi_\nu$ and $\tilde\xi_\mu$ are displayed in Fig.~\ref{fig_Xis}.

The above relations permit to actually calculate the various
coefficients appearing in the previous sections. First, one
finds that $\mathfrak{c}_0 = \nu/2$, so that it suffices to
demand $\nu\geq \frac32$ to ensure self-adjointedness of
the Hamiltonian \eqref{qh0}. As for its semiclassical
counterpart \eqref{HK}, one finds
$$
\mathfrak{K} = \left( \frac{\nu}{2} + \frac{2\mu+1}{4} \right)
\exp\left( \frac{3}{2\mu} \right),
$$
whose minimum value $\mathfrak{K}_\text{min}$
is reached for $\nu=0$ and $\mu_\text{min} = (3+\sqrt{21})/4
\approx 1.89$, at which point one has $\mathfrak{K}_\text{min}\approx 2.64$.

\begin{figure}
\includegraphics[width=0.45\textwidth]{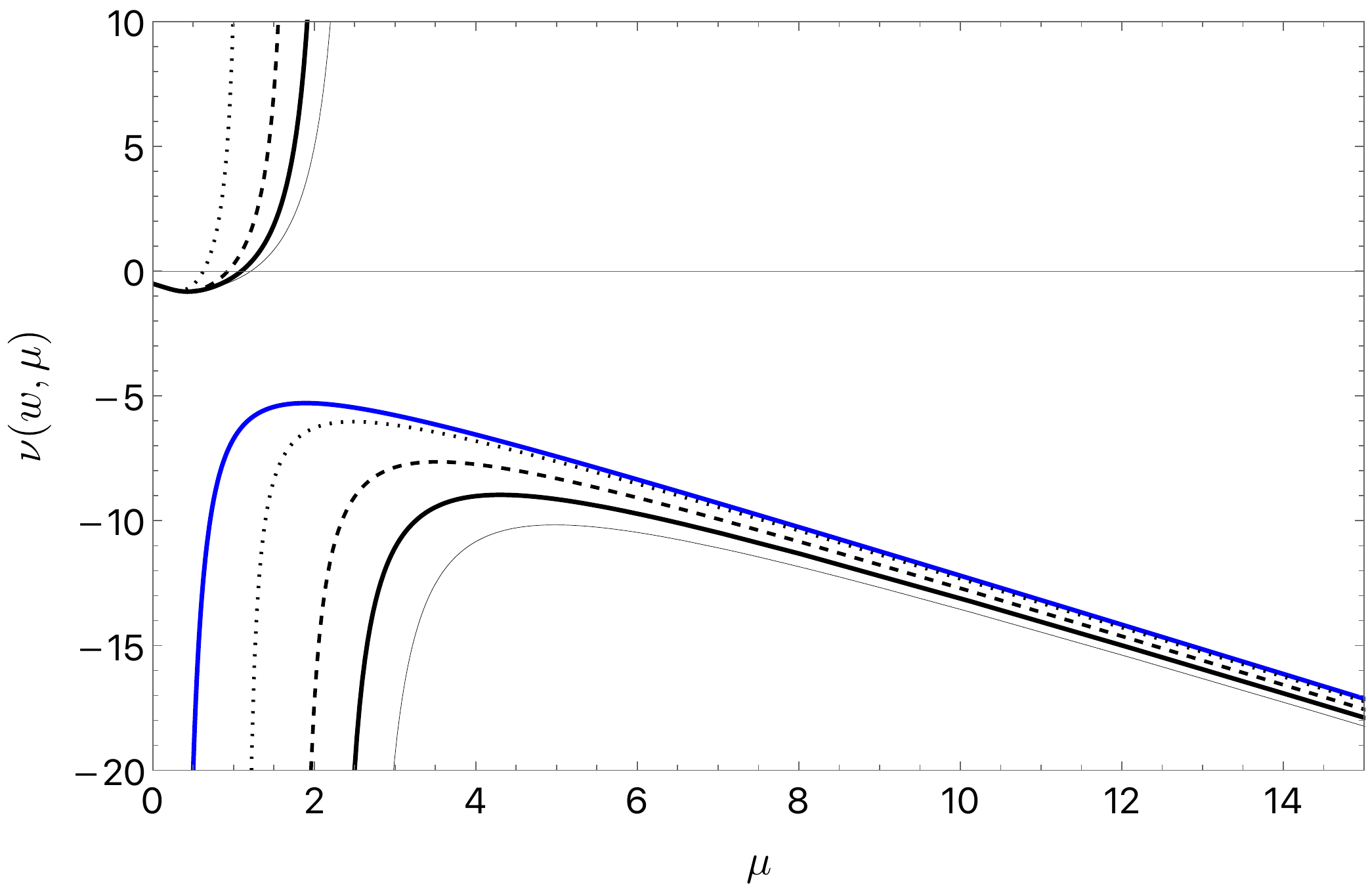}
\caption{Condition \eqref{numuw} on $\nu(w,\mu)$ ensuring the
potential $\mathcal{V}_\MS$ in \eqref{msqvarsigma} takes
the form $(q^{r_2})''/(q^{r_2})$; shown are $\nu(w,\mu)$ for
$w=0$ (thin line), $w=0.1$ (full line), $w=0.2$ (dashed line),
$w=0.3$ (dotted line) and $w=\frac13$ (full blue line).
As both $\mu$ and $\nu$ are positive
definite, it is clear that for a given value of $w$, there is
only a very limited range of $\mu$ satisfying the condition.
For $w=\frac13$, the positive branch disappears and there is
no such solution.
}
\label{fig_NuMuW}
\end{figure}

Moving to the quantum corrections to the evolution of perturbations, we
find 
$$
\frac{\Ts}{\Ns} = \left( \frac14 + \frac{\mu+\nu}{2} \right) \exp
\left[ 
\frac{17-9w}{6\mu (1-w)}
\right],
$$
so that the conformal potential can be cast into the usual $z''/z$ form
if the equation 
\begin{widetext}
$$
\left(\mu+\nu+\frac12 \right) \exp
\left[ 
\frac{17-9w}{6\mu (1-w)}
\right] =
\frac{3(1-w)}{1-3w} \left[ 
\nu + \left( \mu+\frac12\right)
\exp\left( \frac{3}{2\mu} \right)
\right]
$$
has non trivial solutions for $\mu,\nu>0$. This is solved for
$\nu$ as a function of $\mu$ and $w$ through
\begin{equation}
\nu(w,\mu) = \frac{\exp\left( \displaystyle\frac{3}{2\mu} \right)
- \displaystyle\frac{1-3w}{3(1-w)}\exp\left[ 
\frac{17-9w}{6\mu (1-w)}\right]}
{\displaystyle\frac{1-3w}{3(1-w)}\exp\left[ 
\frac{17-9w}{6\mu (1-w)}\right]-1} \, \left(\mu+\frac12 \right).
\label{numuw}
\end{equation}
\end{widetext}
Fig.~\ref{fig_NuMuW} illustrates the behavior of \eqref{numuw} for
various values of $w$. For the conformal radiation case $w=\frac13$,
Eq.~\eqref{numuw} may only be satisfied for $\nu <0$, in
contradiction to the assumption. As expected from the form
\eqref{VMS} of the potential $\mathcal{V}_\MS$, the limit $w=\frac13$
yields an identically vanishing potential, and \eqref{TsNs}
is undefined unless $\mathfrak{K}$ vanishes, which does not
happen with the basis used.

\bibliographystyle{apsrev4-1}
\bibliography{references}
\end{document}